\documentclass[trsc,reprint]{informs3} 
\OneAndAHalfSpacedXI

\usepackage[margin=3cm]{geometry}
\usepackage[utf8]{inputenc}
\usepackage{graphicx}
\usepackage[]{algorithm}
\usepackage{algorithmic}
\usepackage{amssymb}
\usepackage{float} 
\usepackage{longtable}

\usepackage{xcolor}
\usepackage{hyperref}
\definecolor{blush}{rgb}{0.87, 0.36, 0.51}
\hypersetup{
    colorlinks,
    linkcolor={blush},
    citecolor={blue!60!green},
    urlcolor={blue!60!green}
}

\usepackage{natbib}
 \bibpunct[, ]{(}{)}{,}{a}{}{,}%
 \def\bibfont{\small}%
 \def\BIBand{and}%

\makeatletter
\newsavebox{\@brx}
\newcommand{\llangle}[1][]{\savebox{\@brx}{\(\m@th{#1\langle}\)}%
  \mathopen{\copy\@brx\kern-0.5\wd\@brx\usebox{\@brx}}}
\newcommand{\rrangle}[1][]{\savebox{\@brx}{\(\m@th{#1\rangle}\)}%
  \mathclose{\copy\@brx\kern-0.5\wd\@brx\usebox{\@brx}}}
\makeatother     

\newcommand{\ul}[1]{\underline{#1}}
\newcommand{\uul}[1]{\underline{\underline{#1}}}
\newcommand{\alp}{^{(\alpha)}}
\newcommand{\Id}{\uul{\mathbb{I}}}
\newcommand{\Ma}{ {\uul{M}\alp}}
\newcommand{\Na}{ {\uul{N}\alp}}
\newcommand{\Tk}{\mathrm{T}_s}

\newcommand{\AN}[1]{#1}

\TheoremsNumberedThrough     
\EquationsNumberedThrough   
\MANUSCRIPTNO{}

\begin{document}

\TITLE{Parking search in the physical world: Calculating the search time by leveraging physical and graph theoretical methods}
\RUNTITLE{Parking search in the physical world}

\ARTICLEAUTHORS{
\AUTHOR{Nilankur DUTTA}
\AFF{Institut Lumière Matière, CNRS and Université Claude Bernard Lyon 1, \\
         F-69622 Villeurbanne,
         France, \EMAIL{nilankur.dutta@gmail.com}}
         
\AUTHOR{Thibault CHARLOTTIN}
\AFF{Institut Lumière Matière, CNRS and Université Claude Bernard Lyon 1, \\
         F-69622 Villeurbanne,
         France, and \\ \'Ecole nationale des travaux publics de l’\'Etat (ENTPE), Universit\'e de Lyon,\\ F-69518 Vaulx-en-Velin, France}
\AUTHOR{Alexandre NICOLAS}
\AFF{Institut Lumière Matière, CNRS and Université Claude Bernard Lyon 1, \\
         F-69622 Villeurbanne,
         France, \EMAIL{alexandre.nicolas@polytechnique.edu}, \URL{https://www.alexandrenicolas.net}}

}
         
\RUNAUTHOR{Dutta, Charlottin, Nicolas}

\ABSTRACT{
Parking plays a central role in transport policies and has wide-ranging consequences: While the average time spent searching for parking exceeds dozens of hours per driver every year in many Western cities, the associated cruising traffic 
generates major externalities, by emitting pollutants and contributing to congestion. However, the laws governing the parking search time remain
opaque in many regards, which hinders any general understanding of the problem and its
determinants. Here, we frame the problem \AN{of parking search} in a very generic, but \AN{mathematically compact formulation} which puts the focus on the role of the street network and the unequal attractiveness of parking spaces. This problem is solved in two independent ways, valid in any street network and for \AN{a wide range of drivers' behaviours}. Numerically, this is done by means of a computationally efficient and versatile agent-based model. Analytically, we leverage the machinery of Statistical Physics and Graph Theory to derive a generic mean-field relation giving the parking search time as a function of the occupancy of parking spaces; an expression for the latter is obtained in the stationary regime. 
We show that these theoretical results are applicable in toy networks as well as in complex, realistic cases such as the large-scale street network of the city of Lyon, France.
Taken as a whole, these findings clarify the parameters that directly control the search time and provide transport engineers with a quantitative grasp
of the parking problem. Besides, they establish
formal connections between the parking issue in realistic settings and physical problems.
}

\date{}

\maketitle
\HISTORY{}

\section{Introduction}

Cars are designed to move, \AN{but paradoxically} whether or not to use this mode of transportation is oftentimes less a matter of \emph{driving to} a certain destination 
than being able to \emph{park} there. 
Indeed, in many large metropolitan areas, it is hard to overstate the importance
of parking on mobility choices and, more broadly, daily life in a city \citep{shoup2018parking}: The time spent searching for parking exceeds dozens of hours per driver every year (35h to 107h in Western metropolises, according to INRIX survey data \citep{INRIX2017}), at an estimated cost of several hundred Euros a year per driver.
\AN{The burden of parking search is nonetheless unequally shared among drivers and trips: the many trips ended without any cruising contrast with the very long search times experienced in
some cases  \citep{mantouka2021deep}.}
Furthermore, trips that motorists eventually
give up because of the lack of parking space are not uncommon.
(For example, a 2005 survey conducted in 3 French cities indicated that the
percentage of interviewees who had at least once given up a trip (`balking') because of
parking unavailability amounted to 48\% in Grenoble, 67\% in Lyon, and 100\% in Paris \citep{SARECO2005}.) 
Beyond these individual quandaries, cars cruising for parking may represent more than 10\% of the total traffic 
in many large cities (e.g., 15\% in central Stuttgart, 28\% to 45\% in New York), according to surveys and observations of the number of cars driving by a vacant spot before it is occupied  \citep{shoup2018parking,hampshire2018share}; \AN{these cruising cars} contribute to congestion and pollution in city centres
\AN{\citep{gallo2011multilayer}}. 
Admittedly, these quantitative figures are vividly debated and may overestimate the typical share of traffic due to cruising for parking 
by investigating only the most problematic areas \citep{weinberger2020parking}.

Still, there is a broad consensus about the centrality of parking in transport policies. While some cities require developers to add a parking facility to their development plans, through so called minimum parking requirements \citep[chap. 3]{shoup2018parking}, in many others restrictive parking policies have become an essential lever of action for
transport authorities \citep{boujnah2017modelisation,polak1990parking} to curb down the use of private cars, notably as a response to climate change. Besides,
smart cities are on the rise and intend
to alleviate the pains of parking thanks to dynamic parking information and the possibility to book a parking spot, as well as parking guidance systems \citep{al2019smart}.

These points call for more reliable observations of the current state of parking 
in different contexts, but also for a deeper understanding of the process of parking search, so as to be able to predict the impact of hypothetical measures. 
However, for all its importance, the topic of parking is still fairly opaque, with a limited \AN{intuitive} grasp of its main factors and 
their \AN{quantitative} impact. This dim understanding is not surprising given the complexity of the problem, which mingles socio-psychological and economic factors with physical constraints and collective effects. 
Some seemingly natural assumptions about the relation between occupancy
and parking search time \citep{axhausen1994effectiveness,arnott2017cruising} \AN{are commonly used} \citep{geroliminis2015cruising}, but can hardly be reconciled with facts  \AN{(see Sec.~\ref{sub:binomial_approx}).} 


In this paper, the problem \AN{of searching for parking} is framed in a way that is both
\AN{defined in a mathematically concise way} and more versatile than existing agent-based approaches,
\AN{so as to gain quantitative insight into the parking search process in general.} The focus is put on the role of the street network and the unequal attractiveness of parking spaces (due to their location and rates, for instance) among other blind spots of existing approaches (Section 2); some more studied effects, notably the medium-term elasticity of parking demand \AN{(briefly touched on in the conclusion)}, 
are brushed aside. Interesting parallels with problems in the realm of Physics \citep{schadschneider2010stochastic}, more precisely the asymmetric motion of active particles on a graph, then become apparent and clarify some facets of parking search. Importantly, once thus formulated, the parking \AN{search} problem can be solved not only by means of a computationally efficient agent-based algorithm that we developed, but also
by leveraging the powerful machinery of Statistical Physics and Graph Theory  (Section 3).
This leads to analytical formulae for the search time and for the steady-state occupancy. 
\AN{These hold} in remarkably generic settings, for any street network and for a wide range of driving strategies and parking
choices,
\AN{provided that circling is not too intensive.}
These theoretical results are found to be applicable in toy networks as well
as in the complex, large-scale case of the city of Lyon, France (Section 4).
\AN{For sure, some input parameters of the model, first of which 
the drivers' route choices and their perceptions of the attractiveness of parking spots, must be adjusted depending on the context. However, irrespective of these adjustments, our methods are instrumental in advancing the quantitative understanding of parking search determinants beyond case-by-case (city-specific) findings. They also} 
pave the
way for theoretical assessments of e.g. the excess emissions of pollutants due to cruising for parking or the efficiency of smart parking solutions.

\section{Blind Spots of existing Approaches to Parking Search}

\subsection{Literature survey of the determinants of the parking search behaviour}

The behaviours and strategies of drivers in search of parking have been \AN{probed}
by means of field observations (`revealed preferences'), surveys (`stated preferences'), and more recently \AN{virtual experiments} \citep{fulman2020modeling}. \AN{Among} the factors that influence parking behaviour, pricing has been intensely studied and plays a major role in parking choice \citep{brooke2014street,shoup2018parking,gao2021smartphone}, which has prompted dynamic pricing strategies to control the occupancy \citep{chatman2014theory}. The price elasticity of parking volume (i.e., the number of cars that parked in a given duration) generally lies in the range $[-2,-0.2]$, but strongly depends on the context, in particular the average occupancy and the existence of substitutes in terms of parking facilities or mode of transportation (transit service); it is also found to be sensitive to the methodology used, with disagreements between stated
and revealed preferences (see \citep{lehner2019price} for a recent meta-analysis).
\AN{Of course, the relative prices of alternative parking modes, such as car parks, also affect the drivers' decision
to search and possibly cruise for on-street parking or not \citep{assemi2020searching}.
But, importantly, drivers are not always aware of the real parking price and may have a distorted view of the relative on-street and off-street parking rates \citep{lee2017cruising}.}
Another key determinant of parking choice is the location of the parking space, most prominently the distance to destination both for parking on the curb and for off-street parking. There is a marked preference for spots near the ticket machine (if there is one) in parking lots \citep{vo2016micro} or near the facility of interest in rest areas on expressways \citep{tanaka2017analysis} or open parking lots \citep{paidi2022co2}. 
\AN{The premium for on-street parking in terms of accessibility and convenience may partly explain the intensity of cruising \citep{lee2017cruising}.}
\AN{Other \emph{intrinsic}} features of the parking supply
\AN{contribute to parking choices:} the size of the parking space,
the availability of boards with parking guidance information, which may have an influence or not depending on the conditions \citep{axhausen1994effectiveness,tanaka2017analysis}, etc. All these factors (which are to be subsumed into an `attractiveness' variable in the following) thus influence the parking search process.

But \AN{the drivers'} experience during parking search, in turn, can also affect their parking decision. 
\AN{Drivers generally abhor} long travel times, especially in congested conditions \citep{gao2021smartphone}, and queues \citep{tanaka2017analysis}. 
\AN{On the basis of surveys conducted in Brisbane, Australia, \citet{assemi2020searching} found that the purpose of the trip, the arrival time, the parking frequency, and the on-street parking accumulation all affect cruising for parking.
Using GPS tracks collected in the region of Athens, Greece, and exploring the dependencies of the search times with various
machine-learning approaches, \citet{mantouka2021deep} ascertained that  arriving in the morning peak hours and/or making longer trips 
is positively correlated with cruising.
}
\AN{Regarding the search pattern, the} typical driver first drives to the vicinity of the destination (85\% of times \AN{for the participants involved in the parking `game' set by} \citep{fulman2020modeling}, 
where the average on-street occupancy was set to $\phi=99.7\%$). Then, they start circling around it and possibly spiral farther and farther away from it, before eventually quitting the search or heading for an off-street parking lot, after a few minutes \citep{fulman2020modeling} or more \citep{SARECO2005,levy2013exploring,weinberger2020parking,fulman2021approximation,Cerema2015EMD}.
For drivers that keep searching, the time to park is believed to also depend on the
turnover rate of parked vehicles \citep{SARECO2005} and the competition with other cruising cars  \citep{SARECO2005,arnott2017cruising}. \AN{We will nuance this belief below.}

\subsection{Real search times elude common approaches}
\label{sub:binomial_approx}
Despite the interdependence of the search time $T_s$ and parking choices, it is tempting
to try and assess the former on the basis of simple considerations. To this end, and throughout the paper, 
let us consider a situation in which \AN{we know} the parking supply and the global volume
of parking demand \AN{(they are thus held constant)}. Supposing that the average occupancy ${\phi}$ of
parking spots \AN{can be measured}, a basic but nonetheless very common approximation \citep{anderson2004economics,geroliminis2015cruising} assumes that cars move along lanes of spots that are randomly occupied with uniform probability ${\phi}$, and park in the
first available space that they encounter. This is the theoretical foundation of
the \emph{binomial approximation} 
\begin{equation*}
T_s \simeq \frac{T_0}{1-{\phi}},
\end{equation*}
where $T_0$ is the time to drive from one spot to the next one. According to this formula, for ${\phi} \leqslant 99\%$, the mean search time cannot be significantly longer than $100\,T_0$, which is around one minute if spots are adjacent. 
This result is manifestly at odds with the empirical observation of surging search times long before ${\phi}$ reaches 100\% \citep{arnott2017cruising,gu2020macroscopic,weinberger2020parking}; in San Francisco, \citet{weinberger2020parking} even suggest the blocks where cruising cars eventually park might be occupied only at ${\phi}=59\%$. In any event,
 the time to park is drastically underestimated by this approximation and \emph{ad hoc} corrections to mitigate the discrepancy issue \citep{belloche2015street} are devoid of theoretical ground.
 \AN{
\citet{leclercq2017dynamic} proposed a theoretically better grounded extension of the binomial approximation, by positing that drivers cover a distance $l_{ns}$ without no parking spaces, before finding $m$ successive spots. The resulting equation provides better fits to the empirical data, at the expense of two additional parameters which need to be calibrated individually for each area under study and are found to vary widely.
 }
For their part, \citet{arnott2017cruising} rationalised the underestimation of parking search times by
the following factors: 
\begin{enumerate}
\item there are spatial correlations between occupied spots (this `bunching' effect was later addressed analytically by \citet{krapivsky2019simple,krapivsky2020should} and 
\AN{
by \citet{dowling2019modeling} in their non-uniform queuing networks;
}
an approximate expression was proposed by \citet{fulman2021approximation} if the per-block occupancy is known),
\item the occupancy is not constant in time, but undergoes statistical fluctuations, and periods of
higher occupancy have a stronger impact on the mean search time, 
\item the competition between searching cars aggravates the difficulty of the search,
\item circling leads to inefficient double checks.
\end{enumerate}
A simple model with cars moving along a circle strewn with 100 spots and parking in the first available spot was simulated to illustrate these effects. It yielded search times 44\% larger than
the binomial estimate at an occupancy ${\phi} \leqslant 67\%$; unfortunately, the authors could not explain this effect from an analytical standpoint.
\AN{An identical circular geometry and indiscriminate `park-if-vacant' rule had been used in the probabilistic framework developed by \citet{cao2015system}.}
More broadly speaking, it is critical to realise that 
the parking occupancy $\phi$ is a spatio-temporal aggregate, obtained by averaging observations over a geographic area and sampling them in, or averaging them over, a time period.
\AN{(Incidentally, more sophisticated methods have emerged to afford access to the occupancy at higher resolution \citep{yang2017turning})}. Since the relation between occupancy and search time is non-linear, this 
averaging procedure washes away the oversized impact of long cruising times
near hot spots at rush hours.

On the other hand, \citet{weinberger2020parking} insist on the role of the drivers' idiosyncratic behaviours in generating excess travel distances (perhaps they have kept driving because they were arguing about
where to go for dinner or trying to lull a baby to sleep); the excess travel
 may be misconstrued as cruising (e.g., using GPS data) in districts of San Francisco and Ann Arbor where there is actually no lack of vacant spots. For the specific case of San 
Francisco, \citet{millard2020parking} further note that (even in districts with high parking tension) cruising may actually be rare because the scarcity of vacant spots may be internalised in the regular drivers' behaviours and `perceived parking scarcity leads drivers to stop short of their destinations', thus curtailing cruising. However, these caveats do not suppress the ample evidence of cruising for parking in other cities  \citep{SARECO2005,shoup2006cruising,hampshire2018share,INRIX2017} as well as off-street parking lots \citep{paidi2022co2}. 

To settle the debate about the actual parking pain, it is essential to gain insight into what governs the parking search time
and its dependence on the occupancy in realistic settings. The following sections will blaze
a theoretical trail to do so in a rigorous and generic way. But let us first highlight the influence on parking search of oft-overlooked features, in particular the topology of the explored street network and the fact that parking spaces are not equally attractive to drivers.  These effects will be illustrated with particularly simple, somewhat caricatural examples in this section, before delving into the technical complexity of more realistic settings.

\subsection{Topology of the street network}

While previous \emph{theoretical} endeavours have \AN{generally} considered linear or circular geometries for parking lanes \citep{levy2013exploring,cao2015system,krapivsky2019simple,krapivsky2020should,arnott2017cruising},
in reality parking spaces are located on a geometrically more complex network of streets (or alleys in the case of a parking lot), whose topology constrains the motion of the 
cars. Specific street networks have been studied \emph{numerically}, with a more or less
realistic description of their characteristics (see the review in \citep{boujnah2017modelisation}); 
\AN{to single out only one example, \citet{gallo2011multilayer} 
simulated the road network of the city of Benevento, Italy, using a a multi-layer model that decouples the in-vehicle trip
to the destination (with its specific cost for traffic assignment) from the following cruising part.
}
\AN{ \citet{dowling2019modeling} made theoretical progress in the study of street networks with nodes of
non-uniform degrees.}
But the sensitivity of
\AN{parking search times} to the topology has not been investigated.

\begin{figure}[!htb]
    \centering
 
        \includegraphics[width=\textwidth]{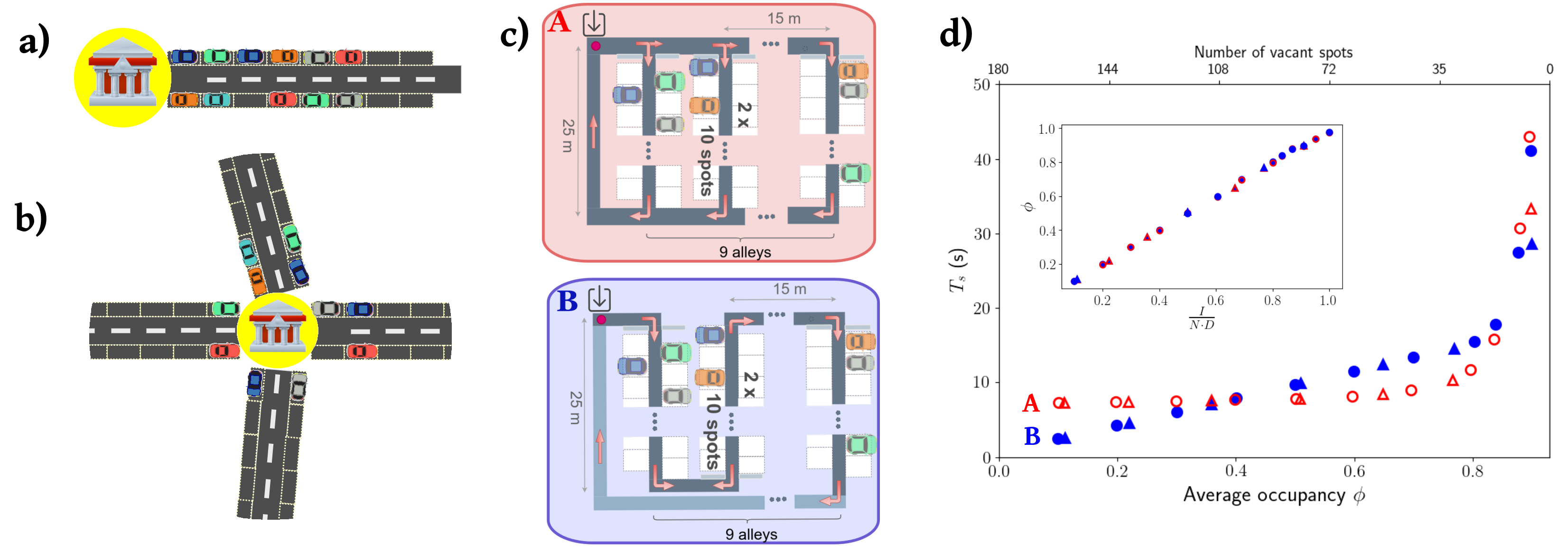}
\caption{Influence of the topology of the street network on (\textbf{a,b}) curbside and  (\textbf{c,d}) off-street parking. A major destination (hot spot) is located at the extremity of a dead-end street in sketch \textbf{(a)} and at an intersection in sketch \textbf{(b)}. Panel \textbf{(c)} displays two model off-street parking lots of $N=180$ spaces; the evolution of the search time $T_s$ in each, for drivers parking in the first vacant spot, is shown in panel \textbf{(d)} as a function of the average occupancy $\phi$.  \emph{Inset}: Relation between $\phi$ and the product of the injection rate $I$ with the parking time $D^{-1}$ [$D^{-1}=30\,\mathrm{min}$ for the triangles ($\triangle$), $60\,\mathrm{min}$ for the circles ($\bigcirc$)]. }
\label{fig:figure1}
\end{figure}

Nevertheless, it is easy to understand that, for an equal number of on-street spots, their 
\AN{spatial distribution and the topology of the street network} will
affect the disutility associated with parking. First consider the simple case
illustrated in Fig.~\ref{fig:figure1}a with a single major destination point (hot spot) and no off-street parking facility. Under equal demand, parking issues will be all the more severe as they are few streets allowing parking on the curb that lead to the hot spot. With fewer incoming streets, it will be harder to find a spot close enough to the destination.

Even if drivers park in the first vacant spot that they encounter, the network topology will matter. To understand this, we turn to the two examples of parking lots displayed
in Fig.~\ref{fig:figure1}b, where cars are injected at constant rate at the entrance and spots
are located at exactly the same positions in space but are not accessible in the same way: 
In topology (A), cars enter one of the 9 parallel rows from the main alley at the top of the sketch, with equal probabilities for each (which is done by setting the turning rates adequately),
whereas in topology (B) cars visit the rows sequentially, driving by every spot before
returning to the entrance point.
At low occupancy, the search time will be mostly constant constant in topology (A), resulting from the travel time along the main alley, 
whereas it will increase almost linearly with the occupancy in topology (B), as the first spots
get more and more occupied. These intuitions are confirmed by the numerical results shown in Fig.~\ref{fig:figure1}c (with the protocol detailed in Section~\ref{sec:implementation}); 
\AN{around $\phi\equiv \phi_c\approx 40\%$, there is a crossover}
from a low-occupancy situation with shorter searches in topology (B) to the opposite case when $\phi>\phi_c$. At high occupancies ($\phi>$80\% or 90\%), the mean search time starts to be dominated by the periods of time when the parking lot is full, because of statistical fluctuations, and cars have to circle several times before a parked car leaves. The search time then surges dramatically and diverges to $\infty$ as $\phi \to$~100\%, with a possibly steeper increase in topology (A), where the rare vacant spot may
be found only after several loops
\AN{(incidentally, this feature will actually elude the mean-field approach proposed below).}
Obviously, the binomial approximation can explain neither
the steepness of this divergence, nor 
the differences between topologies (A) and (B) at moderate occupancies.

\AN{Undoubtedly,} these topological effects can hardly be captured by approaches oblivious of the street network. These considerations \AN{are not} merely abstract: they have contributed to shaping cities and land \citep[intro.]{shoup2018parking}. This is perhaps best exemplified by the strip geometry of strip malls in North America or retail parks in Europe, which enables customers to park directly in front of the shops.

\subsection{Attractiveness of parking spaces}

Beyond the (fixed) structure of the street network, the way in which motorists navigate through it also matters. Variations in the cruising strategies have been observed in parking lots by \citet{paidi2022co2}, but an even more striking example is the anonymous
driver observed by \citet{hampshire2016analysis} who routinely circles around the free parking spaces on the edge of downtown before driving to the paying spaces downtown. Somewhat similarly,
some drivers may tend to circle through the very same blocks over and over again, waiting for a spot to be freed, rather than extending their search to neighbouring streets; the recurring
driving through the same blocks naturally prolongs the search, compared to the exploration of
new blocks. In the same vein, the driver may refrain from parking at the first vacant spot that they encounter and exhibit distinct
preferences, for instance balking at parking too far from the destination or in a paying space when there remains a possibility to
park for free \citep{weinberger2020parking}. 

\begin{figure}[!htb]
    \centering
        \includegraphics[width=\linewidth]{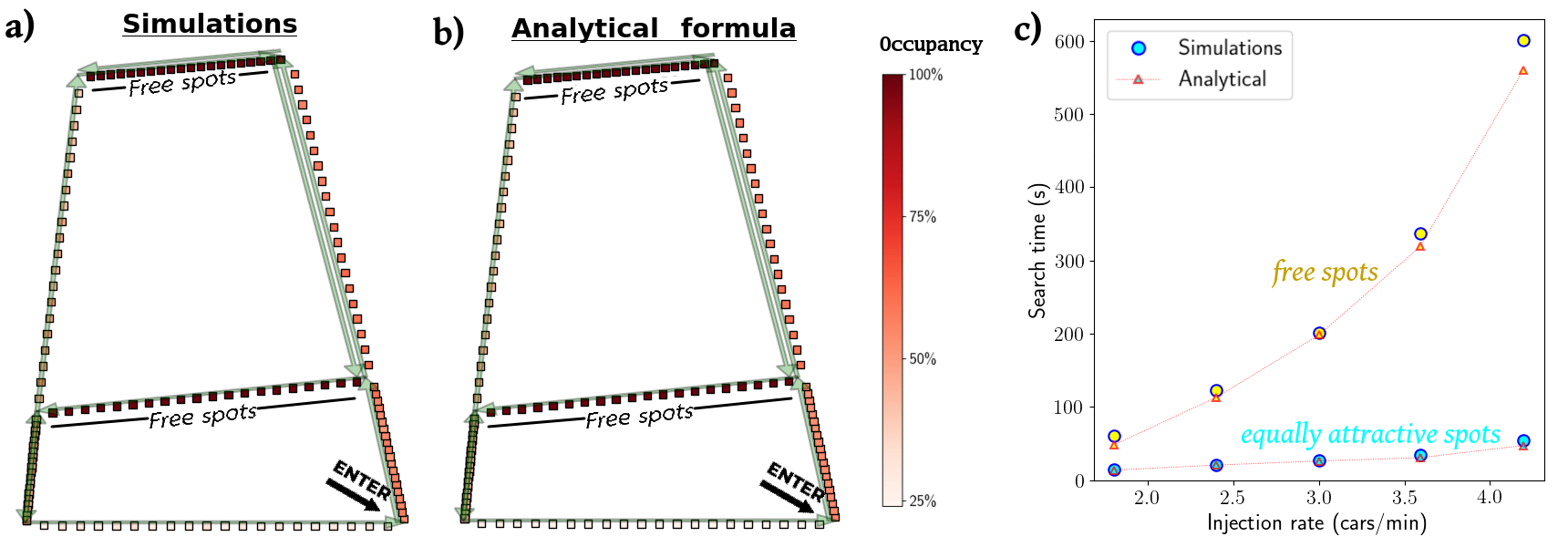}
  
 \caption{Influence of the inhomogeneous attractiveness of parking spots in a small `toy' network. Cars have equal turning probabilities at each intersection. Numerically (\textbf{a}) and analytically (\textbf{b}) derived average occupancy of parking spaces (represented as squares) in the steady state, when drivers are most attracted to free spots ($p_i=100\%$) and only have a $p_i=1\%$ probability to park at every other vacant spot that they drive by. Panel (\textbf{c}) compares the resulting mean search times with the situation in which all spots are equally attractive ($p_i=100\%$). }
\label{fig:smallLyon}
\end{figure}

Quantitatively, we choose to gauge the driver's willingness to park at a given spot $i$ by the probability $p_i$ that she will park there \emph{if} she drives by this spot while it is vacant. The distribution of $p_i$ in a neighbourhood clearly affects parking search. For instance, consider the `toy' network corresponding to a small neighbourhood of Lyon shown in Fig.~\ref{fig:smallLyon}(a); for the same demand, search times will drastically differ between a situation with equally attractive spaces
($p_i=100\%$ for every spot $i$) and a situation in which drivers exhibit a marked preference for two lanes of attractive (e.g., free) spots at $p_i=100\%$ and balk at parking in other (costly) spots, where $p_i=1\%$. In the latter case, drivers are prone to circling before finding a vacant free spot and search times sky-rocket even
though the network-averaged occupancy remains moderate (as also supported by the numerical simulations displayed in Fig.~\ref{fig:smallLyon}(c)).

This simple example shows how using spatially averaged occupancies to estimate parking difficulties goes completely amiss whenever there is a contrast in spot attractiveness: \AN{the} high occupancy of attractive spots
may be balanced by their vacant counterparts. 
\citet{fulman2021approximation} recently improved the binomial approximation by taking into account the occupancies averaged over
small neighbourhoods instead of the global one and demonstrated that it much better reproduces search times in the presence of bunching (an inherent consequence of the sequential parking process) and a spatially heterogeneous demand. Nevertheless, their approach 
requires simulations to estimate of the distribution of local occupancies and does not explicitly handle heterogeneous drivers and parking spots of unequal attractiveness. 
\AN{This does not reflect the spatial and temporal heterogeneity of the parking strategies and cruising behaviours empirically evidenced
by \citet{assemi2020searching}.}

\section{Modelling Framework} 

To take into account the various factors discussed in the previous section, we introduce
a new modelling framework, which rests on an agent-based model with high spatial resolution.
Unlike its forebears, this framework does not postulate specific behavioural rules for the 
drivers looking for parking. Its generic formulation opens the door to a theoretical resolution through the recourse to graph theoretical and statistical physical methods. Besides, it will allow a very efficient numerical implementation.

The model relies on an explicitly described network
of streets (or parking alleys), with parking spots located along them. Motorists are
grouped into distinct categories $\alpha=1,\,2,\,\ldots$ depending on their destination, trip purpose, etc. At an intersection (which is a node of the graph representing the network), they turn into an outgoing street 
with a probability given by the corresponding entry of the category-dependent turn-choice matrix 
${\uul{T}}\alp$. Finally, when driving by a spot $i$, they will choose to 
park there with probability $p_i\alp$ if it is vacant. Parked cars leave their space,
and are thus removed from the simulation, at a rate $D\alp$, which is the reciprocal of the  average parking duration.

Now, in principle, the on-site parking probability $p_i\alp$ depends on a variety of explanatory variables, first of which the parking rate, the distance to destination, and the odds of finding a `better' spot estimated by the driver \citep{levy2013exploring,bonsall2004modelling}, viz.,
\begin{equation}
    p_i\alp= f(\mathrm{rate},\mathrm{distance},...).
\end{equation}

A central idea of the model is to subsume all these variables into two generic variables: (i) an attractiveness $A_i\alp$ reflecting how attractive a spot $i$ is perceived to be \emph{intrinsically}
\AN{(it can estimated with surveys of stated preferences \citep{brooke2016factors,lee2017cruising,antolin2018modelling} or more recent methods based on machine-learning and the distribution of points of interest \citep{nie2021public})}, (ii) the driver's perception of how easy it currently is to park, $\beta\alp \in [0,\infty)$. 
\begin{equation}
    p_i\alp(t)= f(A_i\alp,\beta\alp(t)).
    \label{eq:nonBoltzmann_p}
\end{equation}
At very low occupancy, when parking seems extremely easy, viz., $\beta\to \infty$, the driver will refuse to park anywhere but in their preferred spot, of attractiveness $A\alp_{\max}$ (see examples in the empirical study of \citep{paidi2022co2} for example). 
To the opposite, at extremely high occupancy, $\beta$ will tend to zero and virtually any admissible spot 
($A_i\alp>-\infty$) may be deemed acceptable, viz. $p_i\alp \to 1$.
Since $p_i\alp\in[0,1]$, these considerations invite us to express $p_i$ using a Boltzmann-like functional form, viz.,
\begin{equation}
    p_i\alp(t)= e^{\beta\alp(t)\cdot (A_i\alp-A\alp_{\max})}.
    \label{eq:Boltzmann_p}
\end{equation}

At this stage, we should stress that, in practice, much will depend on the behavioural choices implicitly
encoded in the attractiveness field $A_i\alp$ and the route choices governed by ${\uul{T}}\alp$. Our reasoning holds for any such choices, whether they are realistic (as we
claim for those made in Section~\ref{sec:Lyon}) or not. Therefore, it reaches far beyond current studies focusing on specific rules. In this regard, let us underline that our general formulation encompasses 
the rules used in \AN{many} previous models; \AN{ Appendix~\ref{sec:Appendix_OtherModels} explains how
these specific rules can be transcribed into our formulation.}

\subsection{Simplifications for the current study}

Despite our wish to develop a very general model for parking,
we will make \AN{a few simplifications} in this first, ground-laying work.

First, while in principle each driver evaluates their own $\beta$
depending on their experience along their specific trajectory, here we
assume that $\beta\alp(t)$ only depends on the currently occupied spots in the network.

Secondly, each category of drivers will keep a constant strategy during their search.
In other words, while in reality drivers may modify their turn choices and their perception of spot attractivenesses $A_i\alp$ over time, for instance if they see that their preferred spot is occupied, here the turning rates and attractivenesses will be prescribed once and for all, for each type of drivers. (Note, however, that the actual turn choices and parking decisions may vary due to stochasticity, in  our probabilistic framework, leading to realistic circling behaviour (see Fig.~\ref{fig:cartracks}).

Finally, for simplicity, we do not \AN{simulate} cars after they have left
their parking space; they will simply vanish.
We should also recall that a fixed initial parking demand is considered here. \AN{Should one consider a redesign that affects the parking supply, some feedback of parking search times
on the parking demand should be expected; this point will be succinctly addressed in the conclusion.}

\subsection{Numerical implementation \label{sec:implementation}}
The model is implemented in C++ using data structures that lead to optimal computational performances.
The nodes (i.e., intersections) and edges (i.e., street segments) of the network, and if needed the locations and attractivenesses of spots, are read from CSV files as input and used to create objects of the
Spot, Node, Street, and City classes. 
At each time step ($dt=1\,\mathrm{s}$ in general), a number of car objects set by a Poisson process of 
parameter $I$ (where $I$ is the global injection rate)
are instantiated and injected into the network at one of the entry points. Their positions are measured 
relative to their current street and are updated iteratively at each time step. They switch
streets upon reaching one intersection by randomly selecting an outgoing street according to the specified turning probabilities. We keep track of the (time-ordered) list of spots by which they have driven during the current time step and loop over it to test whether they have chosen to park at any of them, depending
on the attractiveness of the spot and the (constantly updated) parking tension parameter $\beta$. At each time step, another
iteration over currently parked cars removes them with a probability set by their departure rate $D\alp$. 
(The numerical results shown in the previous section were obtained with this implementation.)

Let us finally underline that in our implementation each street is linked to all incoming and outgoing streets and has access to the list of spots located along it; this
allows optimal computational efficiency. More details about the numerical implementation
can be found in Appendix~\ref{sec:app_implementation}; the C++ scripts may be shared to \emph{academic researchers} upon reasonable request to the authors.

\subsection{Mean-field analytical solution for the search time}
\label{sub:ana_method}
Once thus posed, the problem is amenable to theoretical handling. Indeed, having abstracted human behaviour and strategies, we are left with the physical problem of self-propelled particles (namely, cars) moving on 
a directed graph (the street network) and having known probabilities to adsorb at any site (parking spot) along the edges (streets) of this graph. This does not mean in any way that the model describes motorists as brainless particles, but
that these behavioural traits have been encoded into turn choices and attractiveness fields for each type of drivers, which can be processed rigorously.

From now on, it will be more convenient to consider that every street position associated with a parking spot as well as every intersection are nodes of this graph
(notice that the street position where the car starts to park and the parking spot will share the same node label). 
The instantaneous number of cars of category $\alpha$, $\alpha$-cars,
passing by each node \AN{per minute} is then represented by a vector $\ul{I}\alp(t)$ of size $N_{\mathrm{nodes}}$, where $N_{\mathrm{nodes}}$ is the number of nodes and $I_i\alp(t)$ is the number of cars \AN{passing by} node $i$ \AN{per minute} and time $t$ averaged over random realisations.
The drivers' turn choices at the nodes define a transition matrix $\uul{T}\alp$ such that $T_{ij}\alp\in[0,1]$ is the probability that an $\alpha$-car chooses to move from node $i$ to
node $j$ along an edge of the graph in one arbitrary time step, \emph{if} it does not park in the meantime. 
In this graph theoretical approach, $\alpha$-cars initially injected at nodes $j$ (hence, $I_j\alp(t=0)>0$) will be 
\AN{found at} $\ul{I}\alp(t=1)= \ul{I}\alp(0) \cdot \uul{T}\alp$ at the next time step and at 
\begin{equation}
   \ul{I}\alp(K)= \ul{I}\alp(0) \cdot \Big(\uul{T}\alp\Big)^K 
\end{equation}
after $K$ steps, \emph{if they do not park in the mean-time}.
However, at each spot $i$, cars may actually have parked, with a probability $\tilde{p}\alp_i$ given by  $\tilde{p}\alp_i=p_i\alp\,\hat{n}_i$, where $\hat{n}_i=1-n_i$ is zero (one) if the spot is vacant (occupied). Taking this possibility to park into account, the transition matrix $\uul{T}\alp$ should be replaced by $M\alp_{ij}= (1-\tilde{p}\alp_i)\cdot T_{ij}\alp$ and
the spatial distribution
of cars at $t=K$ is actually 
\begin{equation}
   \ul{I}\alp(K)= \ul{I}\alp(0) \cdot  \Big(\Ma\Big)^K.
    \label{eq:Ialp}
\end{equation}

Provided that the occupancy field $(n_i)$ is known, the rate at which an $\alpha$-car 
reaches spot $j$ and parks there reads
\begin{eqnarray}
    P\alp_{j} & = & \sum_{K=0}^{\infty}I\alp_{j}(K)\cdot \tilde{p}\alp_j \nonumber \\
          & = & \ul{I}\alp(0) \cdot \sum_{K=0}^{\infty}  \Big(\Ma\Big)^K \cdot \tilde{p}\alp_j \nonumber\\
           & = & \underbrace{I_i\alp(0) \Big[\Big(\uul{\mathbb{I}} - \Ma\Big)^{-1}\Big]_{ij}}_{R_j\alp} \tilde{p}\alp_j, 
           \label{eq:P_j_alp}
\end{eqnarray}
where Einstein's summation convention (on repeated indices, excluding fixed index $j$ here) is implied and $\uul{\mathbb{I}}$ is the identity matrix. Here, we have considered that the occupancy field $(n_i)$ remains unchanged during the search, which will not hold true in the regime of fierce competition between cruising cars.

Along the same lines, the average `driving, searching, and parking' time $\Tk^{(\alpha,j)}$ of an $\alpha$-car finally parking at spot $j$ (in arbitrary time steps) can be derived; it is the average number of steps $K$ needed to park at spot $j$, weighted by the probability $H\alp_j(K)\cdot \tilde{p}\alp_j$ to reach $j$ after $K$ steps and park there, where 
the total injection rate $I\alp$ of $\alpha$-cars was used to non-dimensionalise $I\alp_j(K)$ into $H\alp_j(K)= I\alp_j(K)/I\alp$. Accordingly, summing over all spots $j$,
\begin{eqnarray}
    \Tk^{(\alpha)} & = &\sum_{K=0}^{\infty}  K\ H_i\alp(0) \cdot \Big[\Ma^K\Big]_{ij}\cdot  \tilde{p}\alp_j \nonumber \\
                     & = & H_i\alp(0) \sum_{K=0}^{\infty}  \sum_{l=0}^{K-1}  \cdot \Big[\Ma^K\Big]_{ij}\cdot  \tilde{p}\alp_j \nonumber \\
                     & = & H_i\alp(0) \cdot \sum_{l=0}^{\infty}   \Big[ \sum_{K=l+1}^{\infty}  \Ma^K \Big]_{ij}\cdot  \tilde{p}\alp_j \nonumber \\
                     & = & H_i\alp(0) \cdot \sum_{l=0}^{\infty}   \Big[ \Ma^{l+1}\cdot 
                     \Big(\uul{\mathbb{I}} - \Ma\Big)^{-1}\Big]_{ij}\cdot  \tilde{p}\alp_j \nonumber \\
                      & = & H_i\alp(0) \cdot  \Big[ \Ma\cdot 
                     \Big(\uul{\mathbb{I}} - \Ma\Big)^{-2}\Big]_{ij}\cdot  \tilde{p}\alp_j.
\label{eq:stime_alpha_spots}
\end{eqnarray}
Equations~\ref{eq:P_j_alp} and \ref{eq:stime_alpha_spots} involve matrices of linear dimension $N_{\mathrm{nodes}}$, which may be very large. However, since each node is connected to a few other nodes at most, these matrices (notably $\Ma$) are particularly sparse. Therefore, their multiplication and inversion can be handled quite efficiently, for instance using the dedicated Python library \AN{(scipy.sparse)}; in particular, we avoid computing the inverse of sparse matrix $\uul{A}\equiv \uul{\mathbb{I}} - \Ma$ and, instead calculate $\ul{Y}\cdot \uul{A}^{-1}$ by solving the linear problem $\ul{X}\cdot \uul{A}=\ul{Y}$.

Incidentally, should an upper bound $K_{\max}$ be set on the number of steps $K$ allowed for parking search before cars quit searching, the foregoing expressions will turn into (see Appendix~\ref{sec:app_capped_search} for the details)
\begin{equation}
   \bar{P}\alp_{j}= P\alp_j - I_i\alp(0)  \Big[\Big(\uul{\mathbb{I}} - \Ma\Big)^{-1} \cdot  \Ma^{K_{\max}+1} \Big]_{ij} \tilde{p}\alp_j
   \label{eq:P_j_alp_capped}
\end{equation}
\begin{equation}
\bar{\Tk}^{(\alpha)} = \Tk^{(\alpha)} - H_i\alp(0)\cdot\left[(\uul{\mathbb{I}}-\Ma)^{-2}\cdot \Ma^{K_{\max}+1}\right]_{ij} \tilde{p}\alp_j,
\end{equation}
where one has arbitrarily defined as  $K_{\max}$ the search time of cars that quit searching. Also note that this gives access to the survival function of the search time, that is to say, the fraction of cars that needed longer than $K_{\max}$ steps to park, which is $\sum_j \Big( P\alp_j - \bar{P}\alp_{j} \Big)/I\alp= 1-\sum_j \bar{P}\alp_{j}/I\alp$.

In all the above formulae, the search time was expressed in arbitrary units, each unit corresponding to the time taken
for a car to travel between two nodes. To recover real time units, we introduce an auxiliary `generating' function
$\uul{N}(z)$ defined by $N_{ij}(z)=z^{\tau_{ij}}\,M_{ij}\alp$, where $z$ is a real variable and $\tau_{ij}$ is the travel time between neighbouring nodes $i$ and $j$ (note that, if $i$ and $j$ are not directly connected, then $M_{ij}=0$). As with the transition matrix, exponentiating $\uul{N}(z)$ into $\uul{N}^K(z)$ gives access to paths of length $K$ steps. This contrivance, detailed in Appendix~\ref{sec:Appendix_searchtime_real}, helps us express the search time in real time units as

\begin{equation}
T_{s}\alp =  H_i\alp(0) \cdot 
\left[(\uul{\mathbb{I}}-\Ma)^{-1}
\cdot N^\prime(z=1)
\cdot (\uul{\mathbb{I}}-\Ma)^{-1}
\right]_{ij}
\tilde{p}\alp_j,
\label{eq:stime_alpha_real}
\end{equation}
where the derivative of $\uul{N}(z)$ satisfies $N_{ij}^\prime(z=1)= \tau_{ij} M\alp_{ij}$

The foregoing formulae were derived for a \emph{given} configuration of the occupancy $\ul{n}$. To get the actual \emph{mean} search time requires averaging over an ensemble of equivalent realisations of $\ul{n}$  (or over time in the steady state). This would be
straightforward if one could compute the average by plainly substituting $\langle n_j \rangle \in [0,1]$ for $n_j =0$ or 1 in the definition of the $M_{ij}$ matrix. Unfortunately, in general, spatio-temporal correlations between spot occupations $n_i$ prohibit such a factorisation.
Still, albeit not perfectly rigorous, this is a valid approximation in the \emph{mean-field} regime, wherein any mutual dependence
between instantaneous spot occupations is neglected. We expect it to be quite reasonable as long
as the search time is not dominated by cars looping over the very same street blocks over and over again, and we thus have an equation (Eq.~\ref{eq:stime_alpha_real}) relating the drive-and-search time to the occupancy field ($n_i$) in this case. Conversely, when circling starts to prevail,
the approximation may lose its accuracy.

\subsection{Analytical solution for the occupancy in the stationary state}

Up to now, it has been assumed that the occupancy of each spot (or its time average) is known. We now purport to show that this occupancy field $(n_j)$ can be derived theoretically, at least in some regimes.
This is achieved by balancing the probability to reach a spot and park there for an incoming car with
the rate of departure of a car parked at this spot.
Generally speaking, the resulting equations will display a dependence on the initial occupation.
In this study, we restrict our attention to the stationary state, where this dependence is washed
away so that methods from steady-state statistical physics are directly applicable. In other words, in the considered time period, the parking demand is assumed to evolve slowly 
enough to strike a balance between incoming $\alpha$-cars and departing ones, viz.,
\begin{equation}
    0= (1-p\alp_{\mathrm{leakage}})I\alp - N \cdot \phi\alp \cdot D\alp,
\end{equation}
where, for $\alpha$-cars, the leakage fraction $p\alp_{\mathrm{leakage}}$ vanishes if all incoming drivers eventually manage to park. Therefore, if only a negligible fraction of drivers quit searching, one arrives at
\begin{equation}
    \phi\alp = \frac{1}{N} \cdot  \frac{ I\alp}{D\alp},
    \label{eq:conservation_alpha}
\end{equation}
and, should all categories of cars have similar parked times $1/D$,
\begin{equation}
      \phi = \frac{1}{N} \cdot  \frac{I}{D},
\label{eq:phi_conservation}
\end{equation}
where $I= \sum_{\alpha} I\alp $.

In addition to this global balance, the rate at which cars park \emph{at any given spot $j$} must be
balanced by the (supposedly constant) departure rate $D$ of parked cars, viz.,

\begin{equation}
      \sum_{\alpha} I\alp P_j\alp = D \langle n_j \rangle.
\end{equation}
Using Eq.~\ref{eq:P_j_alp}, one finally arrives at the mean-field stationary occupancy
\begin{equation}
      \langle n_j \rangle = \frac{ \sum_\alpha R_j\alp p_j\alp I\alp / D }{1+ \sum_\alpha  R_j\alp  p_j\alp  I\alp / D},
\label{eq:nj_stat_self}
\end{equation}
where $R_j\alp$, defined in Eq.~\ref{eq:P_j_alp}, implicitly depends on the $\langle n_i \rangle$'s.

Note in passing that, if there is a single category of drivers, one can express $p_j$ from Eq.~\ref{eq:nj_stat_self} and, on the basis of the empirically
observed occupancies $\langle n_j \rangle$, infer the \emph{empirical attractivenesses} (or more precisely $\beta A_j$) using a fixed-point method. The problem becomes more complex in the more realistic case of multiple categories of drivers and it \AN{should} be addressed in our future works.

\subsection{Numerical validation of the theory on a small street network}

The self-consistent formula, Eq.~\ref{eq:nj_stat_self}, giving the stationary mean-field occupancy is solved with the fixed-point method, using an empty network ($n_j=10^{-5}$) as an initial guess and
iteratively inserting it into Eq.~\ref{eq:nj_stat_self} until the solution converges. 
Applied to the `toy' network of Fig.~\ref{fig:smallLyon}(a), this method yields spot occupancies $n_j$ that
agree very well with the numerical results, as illustrated in Fig.~\ref{fig:smallLyon}(b), with a mean error per spot that does not exceed $\sim5\%$.
Using the analytical stationary occupancy, the search time in seconds is computed with the help of Eq.~\ref{eq:stime_alpha_real}. Figure~\ref{fig:smallLyon}(c) demonstrates the excellent agreement with the simulation results, with the slight exception of the regime of harsh competition at high injection rates.

\AN{
\subsection{Interaction with the through-traffic}

So far, the interactions between cruising cars and the underlying traffic (i.e., the outgoing
and the through traffic) have been overlooked, although cruising can contribute to congestion and generate additional delays. 
Our framework is versatile enough to include this feedback loop with only 
modest changes. Indeed, as Eqs.~\ref{eq:Ialp} and \ref{eq:P_j_alp} are expressed in numbers of steps $K$ and not real time, they hold regardless of traffic-induced delays, because there is no direct competition between cruising cars under our approximations;
the conservation equation, Eq.~\ref{eq:conservation_alpha}, also remains valid, provided that the injection rate is constant. 

On the other hand, the expression for the absolute travel time, Eq.~\ref{eq:stime_alpha_real}, must be amended because the link travel times $\tau_{ij}$ 
now depend on the local density $k_i$. Assume that the underlying traffic is known; it generates an arrival rate of $I^{(0)}_i(k_i)$ cars per hour in front of parking space $i$ (but on the
road). This rate should be added to
the rate of all cruising $\alpha$-cars reaching this site, $R\alp_i$, so that the total flow reads
\begin{equation}
I_i= I^{(0)}_i(k_i) + \sum_{\alpha} R\alp_i.
\label{eq:total_flow_rate}
\end{equation}
Our steady-state assumption implies that this traffic demand can effectively be served, i.e., $I_i$ is lower
than the peak capacity of the street link (otherwise, queues will form and diverge). Accordingly,
we focus on situations with an underlying traffic that does not saturate the system's capacity, even when some cruising traffic is added to it.
We can then use
the fundamental diagram of the road portion to deduce from $I_i$ the local speed, hence the 
actual travel time $\tau_{ij}(I_i)$ from $i$ to $j$, taking due account of the underlying traffic.
Finally, $\tau_{ij}(I_i)$ can be inserted into Eq.~\ref{eq:stime_alpha_real}, where it appears in the
definition of the matrix $N$.


Let us illustrate this method using the `toy' network of Fig.~\ref{fig:smallLyon} (with free spots), for an injection rate of 216 cars in search of parking per hour. For that purpose, we assume a uniform through traffic $I^{(0)}_i=500\,\mathrm{veh/h}$ and
a flow-speed relation, say, $\mathrm{FlowRate}(v)=150\,(v-v^2/v_f)$ (with $v$ in km/h and $v_f=30\,\mathrm{km/h}$).
Using Eq.~\ref{eq:P_j_alp}, we find cruising rates $R_i$ (the $\alpha$ superscript has been dropped
because there is just one category of drivers in that example) ranging from 26 to 240 cruising cars per hour on the different block-faces, on top of the underlying traffic, hence (from Eq.~\ref{eq:total_flow_rate})
total flow rates $I_i$ of 526 to 740 cars/hour. With the proposed flow-speed relation, this results in travel times $\tau_{ij}\propto 1/v$ on the street links 
that are from 1\% to 10\% longer than without cruising traffic. (These values of $\tau_{ij}$ 
can be inserted into Eq.~\ref{eq:stime_alpha_real} to get the total driving times.)

The method could be further refined to account for the slower driving speed of cruising cars, compared to transiting ones, by introducing a dependence of the driving speed $v$, not on the 
total traffic $I_i$, but on \emph{both} the underlying traffic  $I^{(0)}$ and the cruising traffic. 
Although this refinement may be of interest in practical studies, its implementation is straightforward and will not be discussed here.
In fact, in the next section, we shall neglect the contribution of the cruising traffic
to congestion and consider fixed driving speeds on the
street links (e.g., set by the underlying traffic), irrespective of the number of cruising cars.
}

\subsection{Implications}

The theoretical framework exposed in the previous paragraphs is somewhat technical, but it has immediate
practical implications. First, it is easy to understand that, in the mean-field regime, the average parking time $1/D$ only matters relative to the parking demand quantified by the injection rates $I\alp$; in other words, the time unit
of the problem could be reset so that $D=1$. More importantly, these rates $D$ and $I\alp$ do not
directly affect the mean-field parking search time $T_s\alp$ (Eq.~\ref{eq:stime_alpha_real}); their
effect is mediated by the occupancy field $\underline{n}$ and the average occupancy $\phi$, 
\AN{consistently with \citep{dowling2019modeling}}.
This notably implies that the turnover rate, although widely believed to be central for the parking tension, only impacts the parking search process via its influence on the average occupancy. Put differently, as long as the fraction of time during which spots are occupied remains constant in a homogeneous period, increasing the turnover rate does not reduce the search time.  Of course, in practice, limiting the parking time by rule or by cost \emph{will} ease the parking pain, by altering the parking demand and the average occupancy -- but not \emph{per se}.

One should however bear in mind that these results are rigorous only under the mean-field hypothesis, which notably
breaks down when drivers start circling significantly and competing for freshly vacated spots. This breakdown is particularly conspicuous in the high-occupancy (i.e., rightmost) part of Fig.~\ref{fig:figure1}(d) (where there is no longer a unique relation between the search time and the occupancy, especially for Parking Lot A).

\section{Large-scale Application of the Method: On-street Parking in Lyon, France \label{sec:Lyon}}

So far, we have shown that our theoretical and numerical modelling framework is
applicable in small-scale street networks, but its scalability to
the larger-scale networks of actual cities has not been proven yet. This section is aimed at demonstrating that
our methods are quite efficient in rendering parking search in a large city. It will \AN{also expose} how the model can be
used in practice and what input data are necessary to this end.
This will be illustrated with the morning peak hour (7am to 10 am) in the city of Lyon, as of 2019, which belongs to the second
urban area in France, with a municipal population around 500,000 people, and features
well-known difficulties to park in its centre \citep{SARECO2005}. For instance, according to 2015 household travel surveys \citep{Cerema2015EMD}, 55\% of the respondents with a place of work or study in Lyon mentioned parking issues and the percentage peaked at 60\%-70\% for those who would not drive all the way to the place of work or study. That being said, let us make it clear that our aim is not to provide the most accurate picture of parking in Lyon, which would require more input data than we own,
but rather to put to the test our methods in a fairly realistic case study.

\begin{figure}[!ht]
    \centering
   \includegraphics[width=\linewidth]{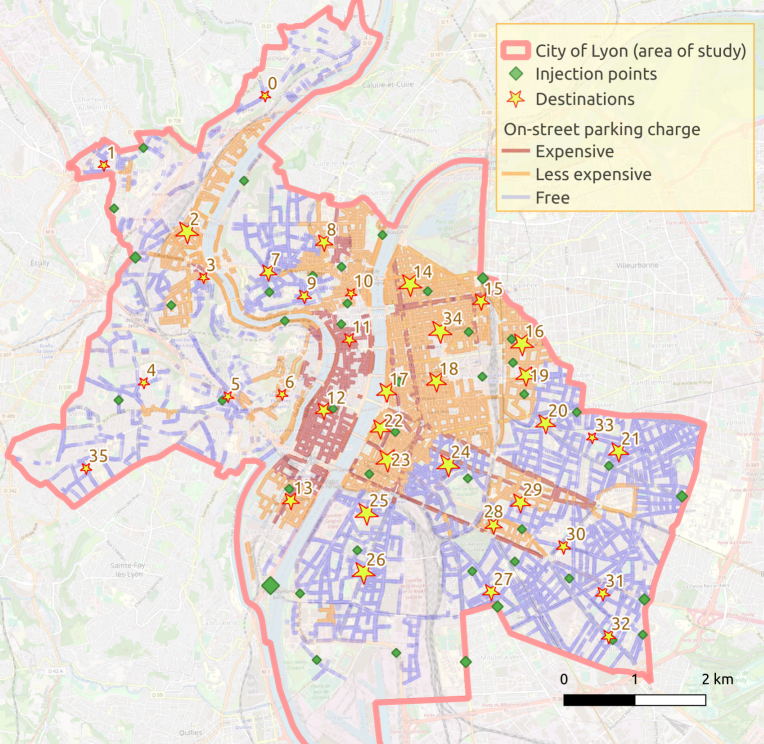}
        \caption{Map of the city of Lyon with its 84k on-street parking spots (as of 2019) and
        the injection points and destinations implemented in our model (the symbol sizes
        represent the injection rates of cars associated with these points). }
        \label{fig:mapLyon}
\end{figure}

In particular, we resort to crude Origin-Destination matrices for Lyon-bound cars in the morning, on the basis of the number of people living and working in each district (`\emph{arrondissement}'), corrected by a plainly empirical factor to reflect the proportion of drivers trying to park on the curb. Note that more accurate matrices based on GPS tracking are commercially available, but expensive. More precisely, 46 injection points (`sources') are chosen in, and at the boundary of, the city. For entry points inside the boundaries, the relative rates at which cars are injected at these points
depend on the population of the \emph{arrondissement}. For points located on the boundary, they depend on the estimated inflow of cars from outside the city. The latter points account for about half of the injected cars. The global injection rate
will be varied in the following. Regarding the destinations, 36 points
were selected (Fig.~\ref{fig:mapLyon}) and attract a fraction of the injected cars that is
roughly proportional to the local number of jobs, upon aggregation \AN{by} \emph{arrondissement}; empirical correction factors intended to reflect the availability of private and off-street parking were introduced manually (Table~\ref{tab:SI_destinations}). We assume \AN{that the injection points and the destinations are decoupled}: cars are randomly bound to a destination, with the same probabilities irrespective of where it was injected. 
Also note that, even though our theoretical framework can include off-street parking, only drivers looking for on-street parking are considered here; parking search will therefore be described as unsuccessful if the drivers eventually opt for off-street parking or if they give up their trip altogether.

\begin{figure}[!ht]
    \centering
   \includegraphics[width=\linewidth]{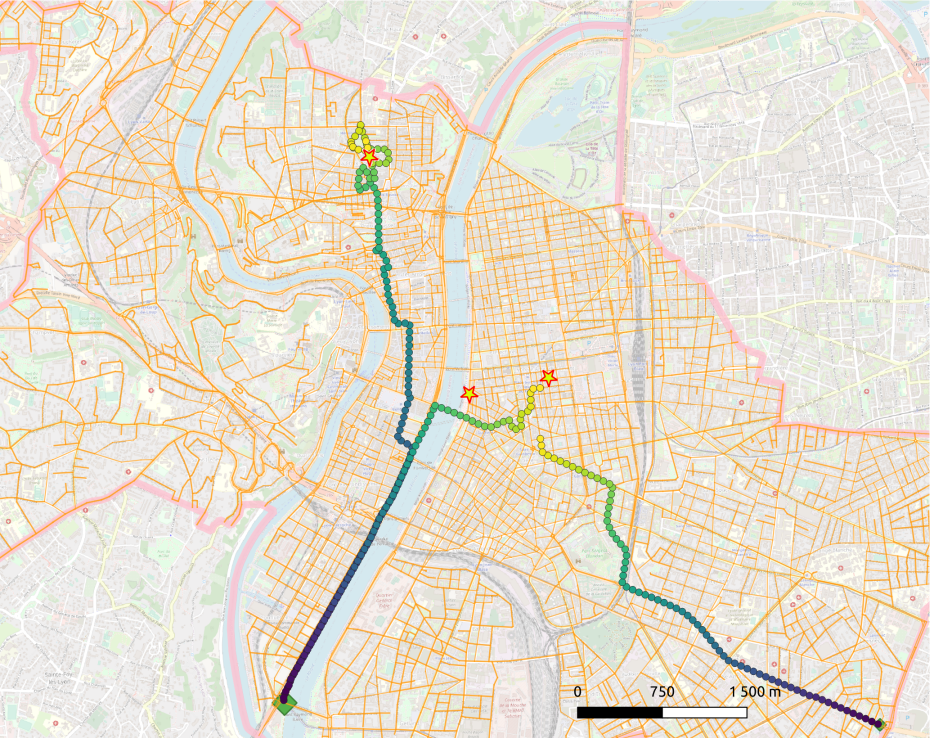}
        \caption{Examples of a few typical simulated car trajectories, from the injection point
        to the parking spot near the destination, represented by a star, for a global injection rate of 50 cars/min. The colours inside the circles denote the time passed since the car's entry in the network. }
        \label{fig:cartracks}
\end{figure}

Detailed information about the locations of the $\sim$ 84,000 on-street parking spots, their rate, and their occupancy (as of 2019) as well as the street network was provided to us by the City of Lyon.
Here, their attractiveness is assumed to depend exclusively on the Euclidean distance
to the driver's destination and on the parking rate (either free of charge, or 1 Euro per hour or 2 Euros per hour). The hourly rate $c_h$ in Euros is converted into an equivalent additional distance to destination $d_\mathrm{charge}\approx c_h\cdot 200\,\mathrm{m}$ by balancing the cost for a dwell time of 3 hours with the cost of walking a distance $2\cdot d_\mathrm{charge}$ (from the spot to the destination and the other way round), under the premise of a value of time around 13 Euros/h, multiplied by a penalty factor of 2 for the effort of walking \citep{bonsall2004modelling}, and an effective walking speed of $1\,\mathrm{m/s}$ (with respect to Euclidean distances). Finally, we consider that the attractiveness decreases with the square of the distance to destination, with a characteristic length $d_\mathrm{walk}=250\,\mathrm{m}$, so that
\begin{equation}
A_i= -\frac{{d_\mathrm{dest}^{(i)}}^2+(200\,c_h)^2}{d_\mathrm{walk}^2}.
\end{equation}
 The specific expression that we chose could be refined with targeted local phone surveys, field studies, or `virtual experiments' \citep{fulman2020modeling} involving a representative panel of participants.
 
 Regarding the perceived parking tension, we \AN{arbitrarily} adopt the following expression for the dependence of the tension factor on the instantaneous local occupancy
 \AN{
 ${\phi\alp}$ (measured within a disk of radius $d_{\mathrm{walk}}$ around the destination), $\beta\alp=\frac{1-{\phi\alp}}{{\phi\alp}} + \epsilon$,
 where $\epsilon \ll 1$ (in practice, $\epsilon$ is set to 0.1) is introduced to prevent the
 selection of too distant spots at high occupancy.
This assumes that the actual local occupancy close to the destination reflects the 
parking tension experienced by each category of users.
 }
 It is noteworthy that the formula verifies $\beta\alp \to \epsilon \approx 0$ at very high occupancy and $\beta\alp \gg 1$ when $\phi\alp$ is small.

However, it should be realised that a large fraction of the existing spots will remain occupied over the whole simulation period (i.e., the morning peak hour). Estimating \AN{(in light of limited field data)} that 95\% of spots in the centre and 87\% in the periphery are occupied at a given time in the morning rush hour, this long-term occupancy was replicated by randomly
declaring forbidden (`frozen') a corresponding fraction of spots, \AN{reduced by 25\% to 30\% 
to account for cars departing during the simulation period.} Besides, the mean parking time was set to 2.5 hours, for all destinations (note that, in any event, it
cannot exceed the duration of the simulated period).

Figure~\ref{fig:mapLyon} gives an overview of the locations and rates of parking spots, as well as the chosen injection points and destinations.
\AN{
Overall,} cars are directed towards their target by computing the shortest-path distance $\tilde{d}$ of every node at the end of a street segment, with the help of the Dijkstra algorithm, and 
favouring turns into streets whose ends are closest to the destination.
\AN{However, deviations from this shortest-path are allowed, all the more so as the car gets closer to its destination.
} 
Technically, 
for every category $\alpha$ of drivers (i.e., destination), the probability of turning 
into an adjacent street $\mathcal{S}_i$ when a car in street $\mathcal{S}_0$ reaches an intersection is given by
\begin{equation}
    T\alp_{\mathcal{S}_0 \to \mathcal{S}_i} = \frac{1}{Z} e^{\eta[ \tilde{d}\alp(\mathcal{S}_0)]\cdot\frac{\tilde{d}\alp(\mathcal{S}_0)-\tilde{d}\alp(\mathcal{S}_i)}{l_i}}, 
\end{equation}
where $l_i$ is the length of street segment $\mathcal{S}_i$, $Z$ is a normalising factor
that makes the turning probabilities to adjacent streets sum to one, and the coefficient
$\eta(\tilde{d})=\min(5,\,\frac{\tilde{d}}{500\,\mathrm{m}})$ results in more deterministic trajectories 
far away from the destination and more fluctuations when approaching it or while cruising, where circling or spiralling behaviours are expected. We do not explicitly model the interactions with transit-related traffic and assume that cars have a constant speed $v\approx 22\,\mathrm{km/h}$ throughout the city in the morning peak hour.
Typical examples of car trajectories to their parking spot, simulated with these turning probabilities, are presented in Fig.~\ref{fig:cartracks}. One can see that the route choices to the target destination look quite reasonable and that they are  possibly followed by circling in the vicinity of the destination if no attractive spot is found in the first place.

\begin{figure}[!htb]
    \centering
 
        \includegraphics[width=0.62\textwidth]{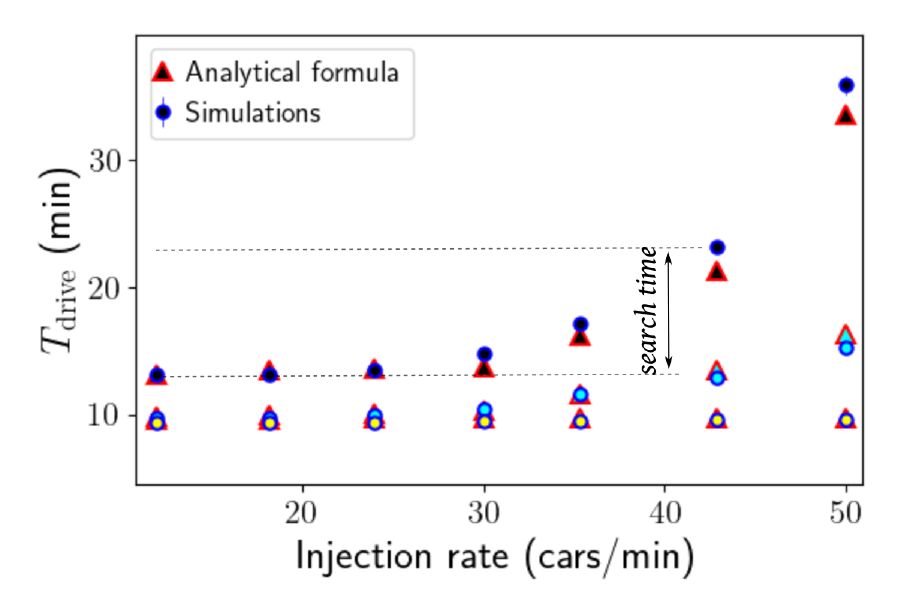}
\caption{Dependence of the mean total driving time, including the curbside parking search time, for two categories of drivers (i.e., two destinations, irrespective of the entry point) on the global injection rate. Black-filled symbols represent destination 10 (C{\oe}ur Croix Rousse), whereas cyan-filled and yellow-filled symbols represent destination 12 (Ainay) and destination 18 (Part-Dieu), respectively.}
\label{fig:searchtimes_Lyon}
\end{figure}

\begin{figure}[!htb]
    \centering
 
        \includegraphics[width=0.7\textwidth]{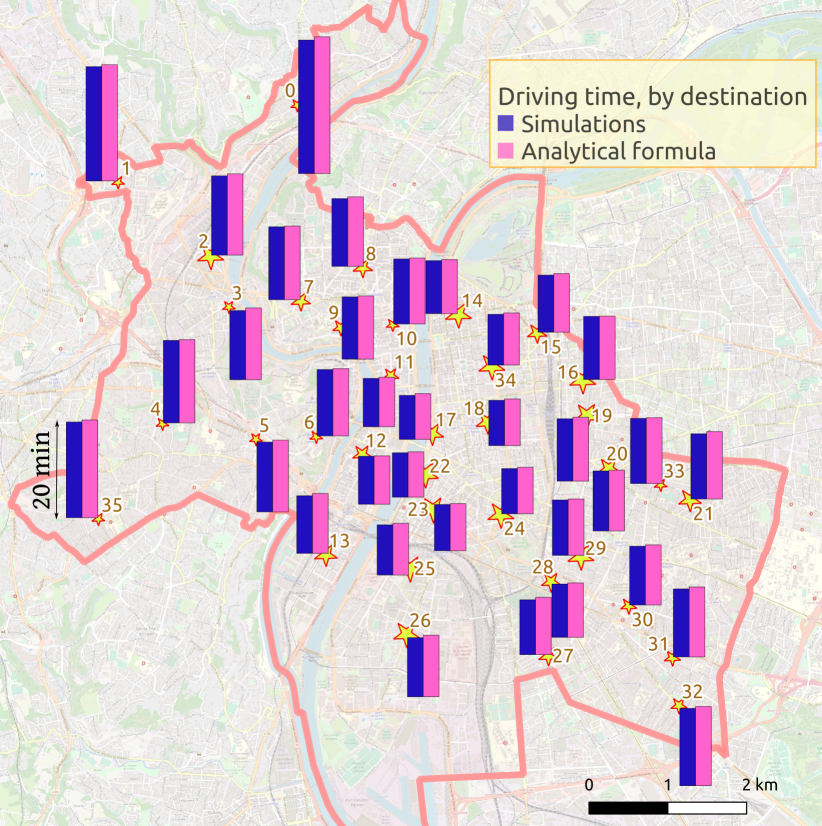}
\caption{Map of Lyon comparing the the simulated mean total driving time and its analytically derived counterpart for all possible destinations (drivers' category), irrespective of the entry point, for a global injection rate of 24 cars/min.}
\label{fig:Lyon_stimes_dest}
\end{figure}

Our highly efficient computational methods enable us to simulate the driving and parking of 
$\sim 10^4$ cars in the whole city (containing $\sim 10^4$ street blocks)  over a period of 3 hours in a matter of tens of seconds, using a single CPU core on a personal laptop. The outcome
of the simulations in terms of average travel times (including parking search) is shown in Fig.~\ref{fig:searchtimes_Lyon}, as a function of the global injection rate, and distinguished between destinations in Fig.~\ref{fig:Lyon_stimes_dest}.  The excess travel time due to parking search is the difference
between the observed travel time and that found for vanishing demand (injection rate). Defining the parking search time
in this way circumvents the ambiguous definition of the start of the parking search or the onset of cruising, which was pointed out in \citep{millard2020parking}.

Figure~\ref{fig:searchtimes_Lyon} shows that the mean search time mostly lies between a few tens of seconds and $\sim 5$ minutes at experimentally reasonable injection rates. 
\AN{Note that the search times start to surge, although the average occupancy $\phi$ \emph{over the whole city} still lies significantly
below 80\%.}
\AN{To test these results against empirical data,} we have analysed the results of a massive travel survey conducted in 2015 \citep{Cerema2015EMD} 
\AN{in which the respondents were asked about their parking search times, among other questions. We have} arrived at a mean reported search time  between 2 and 3 minutes for people who parked on the curb in Lyon in the morning peak hour (the survey data contain 364 such trips), which is broadly consistent with the foregoing figures.
However, we must plainly admit that we lack empirical data to rigorously gauge the accuracy of the numerical results. At best, we can say that the average search times in the simulations are reasonable and that the total travel times reported in Fig.~\ref{fig:Lyon_stimes_dest} are compatible with the
average time spent in trips in a day, which lies around 70 minutes in the Lyon area \citep{Scot2018}, but 
this compatibility is more a safety check than a stringent validation. Furthermore, the output of the model in terms of spot occupancy is of course realistic, but this cannot be used as a touchstone, because empirical occupancy data were used as input. Lastly, the search times are found to be strongly dependent on the destination and on the injection rate, which makes sense but also urges to take with a grain of salt any empirical validation of a model based on data from a single day or place, especially if the parking demand was adjusted freely. That being said, it must not be forgotten that 
our goal is not to fine-tune the calibration of the model parameters to reproduce the conditions of a specific day in Lyon as closely as possible, but instead to validate the formal connections that we have established between the occupancy and the search time, as a function of the injection rate and the drivers' preferences.

Indeed, we can now make use of the theoretical \AN{arsenal} introduced above to predict travel times
without resorting to extensive simulations. 
\AN{
Starting with a random initial guess for the occupancy field $n_j$
the parking tension factors $\beta\alp$ are calculated based on the average local occupancies $\phi\alp$rmse and then serve as input for the fixed-point equation
giving the occupancy field $n_j$, Eq.~\ref{eq:nj_stat_self}; the process is iterated until convergence.
}
This occupancy field, in turn, is used to calculate the average driving time (including parking search) via Eq.~\ref{eq:stime_alpha_real}.
Figure~\ref{fig:searchtimes_Lyon} underscores the quality of these analytical predictions in the regime of low to moderate competition for spots.
The analytical results also display concordance with the numerical ones if the travel times of cars are inspected separately for every destination, as shown in Fig.~\ref{fig:Lyon_stimes_dest} for an injection rate of 24 cars looking for on-street parking per minute: the root-mean-square relative error on these times is smaller than 3\%, while the root-mean-square absolute error on the occupancy is under 0.04 in that case. Given the complexity of the street network, the possibly intricate car trajectories (see Fig.~\ref{fig:cartracks}), and the multiple car categories and attractiveness fields, the level of agreement displayed in Fig.~\ref{fig:Lyon_stimes_dest} is remarkable. We should stress that these results were obtained using \emph{only} the above analytical formulae, and not the output of the agent-based simulations; even the parking tension parameters $\beta\alp$ were determined theoretically.

\AN{
Of course, besides the search times, these calculations also give access to a variety of other indicators. For instance,
the total parking revenue can readily be obtained
by multiplying the parking rate by the average occupancy per spot. Using the self-consistent analytical expression (Eq.~\ref{eq:nj_stat_self}) for the occupation number, the parking revenue generated by cars parking on the curb in Lyon in the morning peak hours is estimated at 4,720~Euros per hour (for a global injection rate of 43 cars per minute); this figure is close to the value of 4,666~Euros per hour obtained by direct numerical simulations of the model.
}

\section{Conclusions}

To summarise, our study of the parking search process has unveiled a relation between the total travel time and the occupancy of parking spaces, which takes into account the effect of oft-overlooked factors such as the particular topology of the street network and the \AN{heterogeneous} attractiveness of parking spaces. 
\AN{The major contribution of this paper is to quantitatively account for these effects, inherent to the problem of
parking search, by propounding an algorithm and equations of broad applicability.}
The derived equations are \AN{indeed} generic and can therefore be applied to \AN{a wide range of} parking search strategies and to any street network in the world, from simple parking lot models to large cities.

While it lays the ground for a rigorous approach to the topic,
the present work involves some simplifications that limit its capacity to describe the full scope of parking-related issues. Firstly, the framework introduced here discards possible changes in behaviour of the drivers during their search (e.g., if they have a specific favourite spot and realise that it is occupied) and it handles parking decisions spot by spot, in a sequential way, overlooking the possibility to see other spots at a distance; the drivers' perception of parking tension is also handled as a function of the occupancy of the network, instead of depending on the drivers' observations. 
Secondly, the generic mean-field formula that we derived for the search time requires detailed information about the occupation of spots in the network; we showed how to theoretically compute such information, but only 
\AN{when circling remains moderate} and in the stationary regime. 
\AN{The meaning and implications of the stationary assumption have been discussed previously; see Dowling et al. \cite{dowling2019modeling}.)}


Still, the parallels that have been drawn with well posed physical problems (namely, the motion on a directed graph of self-propelled particles that can adsorb on its edges) bear appealing promises: the foregoing limitations may indeed be overcome by tackling more refined physical models, which may incidentally open new avenues of research for physicists and mathematicians. Even more importantly, these abstract connections can clarify the influence of some parking 
characteristics on the occupancy and the search time (for instance, the parking time and the turnover rate, in this contribution).

To conclude, we would like to highlight the potential of this theoretical approach
to address practically relevant issues that may be out of reach of agent-based simulations.
While this study revolved around the description of a reference scenario, the formulae giving the parking search time
can be integrated into a static multi-modal transport model, as part of the generalised cost of a motorised trip, in order to evaluate hypothetical scenarios of urban redesign.
This would notably account for the elasticity of parking demand to search times and enable one to assess whether, for instance, reducing the parking supply increases the emission of pollutants by generating more cruising traffic or acts in the opposite way because scarcer spots and longer search times deter travelers from using their cars.

The importance of our work is even clearer if one tries to optimise some urban redesigns, which typically require numerous simulations that can be bypassed by our analytical formulae. Let us mention a few examples. One may naturally wish to restrict the parking supply while holding the cruising traffic in check. But one may also turn to adaptive parking pricing, whereby the attractiveness of spots is modulated by changing the parking rates and which has already been experimented in San Francisco and Los Angeles, notably \citep{shoup2018parking}. These modulations affect not only the demand, but also the parking search time \citep{dutta2021searching}; our formulae relating the search time to the attractiveness in the stationary regime can help find an optimal spatial modulation of the parking rates in busy neighbourhoods.
Finally, in the context of the development of smart cities and the emergence of smart guidance and parking reservation system, our work paves the way for an assessment of the maximal performances that can be expected from smart guidance applications. This will be the focus of a forthcoming manuscript.

\ACKNOWLEDGMENT{
We thank Y. Pachot, C. Marolleau, and N. Keller-Mayaud from the City Council services of Lyon for giving us access to empirical data about parking, and L\'eo Bulckaen for enlightening discussions. This work was funded by an Impulsion grant from IDEXLYON (2020-2021).
}

\begin{APPENDICES}

\setcounter{equation}{0}
\renewcommand{\theequation}{S\arabic{equation}}
\setcounter{table}{0}
\renewcommand{\thetable}{S\arabic{table}}
\setcounter{figure}{0}
\renewcommand{\thefigure}{S\arabic{figure}}%

\AN{
\section{How other agent-based models are encompassed by the proposed framework}
\label{sec:Appendix_OtherModels}

There already exist a number of parking microsimulation models. In the main text, we have claimed that the proposed modelling framework, albeit mathematically concise, can encompass the gist of many of these models, as far as the parking search process is concerned (i.e., without considering the feedback loop of parking search on the parking demand, which pertains to the higher-level transport model into which our parking model can be integrated). This section supports this statement.

Let us start with the recent PARKGRID model described in \cite{fulman2021approximation}, in which
agents are injected into a grid-like city with a Poisson rate of injection (as in our case). 
They start searching a parking space within 500~m of their destination, driving along street links and parking in the first available space, regardless of the parking rate. In our framework, this
means that all spots within 500~m of the destination have the same attractiveness $A_i=0$ and the other ones have an attractiveness $A_i \to - \infty$. If drivers reach the end of the 
street link before finding a vacant spot, they will turn to an adjacent link with probabilities set according to the results
of earlier virtual experiments similar to those in \cite{fulman2020modeling}. These probabilities
can straightforwardly be encoded in our matrix of turning probabilities $\uul{T}$. Finally,
if they fail to park after 20 minutes, they stop searching for on-street parking, which
is equivalent to setting our upper bound for the parking search time to 20 minutes. Accordingly,
the foregoing rules can naturally be described within our framework. Note, however, that this
paper assumes a uniform distribution of parking times, whereas we considered a constant departure rate here.

Turning to PARKAGENT \citep{benenson2008parkagent}, the route choice is simpler in this model, insofar as drivers choose the street segment that brings them closest to their destination, which
can of course also be encoded in our turn-choice matrix. 
They start searching for parking when they are within a distance of 100~m (rather than 500~m) of the destination and reduce their driving speed to do so, which can be implemented in our model (because driving speeds on street links may vary with the category of drivers), although we have not done it in our test studies. But interestingly they do not necessarily park in the first vacant spot. Instead, they assess the odds of finding a vacant spot before the destination, on the basis of the percentage $p_\mathrm{unocc}$ of vacant spots among those that they have seen so far and the estimated number $N_s$ of 
spots to be encountered before reaching the destination. Taking this into account, they will
accept a spot with probability $\max[0, \min(1, 1.5-0.5*p_\mathrm{unocc}\cdot N_s)]$. In our model, this can approximately be transcribed by setting $\beta\alp=1-\phi \approx p_\mathrm{unocc}$ (where $\phi$ is the occupancy) and $A_i$ equal to 1.5 times the negative of the distance from spot $i$ to the destination, measured in number of spots, if one accepts that  $e^{-1.5x}$ is of order 1  for $x<1$ and much smaller than 1 for $x>3$. Nevertheless, note that in this formula the percentage of vacant spots is assessed based on the
actual occupancy $\phi$, and not the user's estimate $p_\mathrm{unocc}$, which we cannot transcribe
in our model. 
If drivers have failed to park 
before reaching their destination, they  will park in the first vacant spot within a disk whose radius grows with time. The new strategy is tantamount to setting $A_i$ to $0$ or $-\infty$ depending on the distance from spot $i$ to the destination, as in PARKGRID, but so far we
have not implemented the possibility of changes in parking strategies over time. Still, 
the change in strategy should be of little importance in the simulations, because the specified route choices appear to maintain the driver close to the destination, where the probability to 
accept a vacant spot is very high.
Finally, drivers consider the search unsuccessful after 10~minutes, rather than the aforementioned 20~minutes in PARKGRID.

The SUSTAPARK model \cite{dieussaert2009sustapark} hinges on a discrete-choice model (with 
logit probabilities) for the choice of a parking mode (on-street, private, or garage) and also
for the probability $p_i$ to accept a vacant on-street parking space. The utility $U_i$ of the latter
is a linear combination intrinsic characteristics of the spot, to which a term describing the  occupancy $\phi_{\mathrm{street}}$ in the street is added. The resulting parking acceptance probability $p_i=\frac{1}{1+e^{-\beta_{\phi}\phi_{\mathrm{street}}}e^{-U_i}}$ can be recast into
the general form of Eq.~\ref{eq:nonBoltzmann_p}, but not
into the Boltzmann-like form of Eq.~\ref{eq:Boltzmann_p}. However, the methodology exposed in Sec.~\ref{sub:ana_method} does not rely on this functional form and is therefore still applicable.
On the other hand, our model was only applied to on-street parking; it does not
describe the changes in parking mode implemented in SUSTAPARK.
Finally, drivers follow random routes close to their destination, which can be encoded in our
turn-choice matrix, but drivers change their speed when they start searching for parking.

Axhausen et al.'s microsimulation introduced in \cite{horni2013agent} in view of an integration in MATSim also considers different parking strategies, but it presents specific features for
the search process. First, agents may either drive directly to the destination and start
searching from there, or start their search a given distance away from it. An easy way to
take this into account in our analytical derivation is to consider that the agents are injected
at the position where they start searching for parking. Drivers then accept a parking space
depending on its distance to destination (as in our model), but also on the elapsed search time.
This specific feature, which introduces a dependence on the driver's history, cannot be accurately replicated by our approach. Still, on a more \emph{qualitative} level, our parking tension parameter
$\beta$ has the same effect as the elapsed search time: parking acceptance will be
enhanced by both of these variables. Regarding route choices, they are determined probabilistically once the search has begun, which is compatible with our approach.

Finally, the approach most closely related to our model is probability the network of finite-capacity queues introduced by \citet{ratliff2016observe}, \citet{dowling2019modeling}, and \citet{tavafoghi2019queuing}, in that
it also involves analytical developments; this elegant approach was employed to address the problem
of estimating the cruising traffic on the basis of empirical occupancy data \citep{dowling2019modeling}. In the model, 
street segments are block-faces on a Manhattan-city grid and each one of them is handled
as a finite-capacity queue that serves the driver if it contains a vacant space, or rejects it 
to adjacent block-faces otherwise. We note
that, compared to ours, this model rests on the following additional assumptions: \\
(i) drivers at junction select the next street segment purely randomly and the nodes of the street network only differ by their degree,\\
(ii) every agent will accept the first vacant space they encounter (which translates into a uniform attractiveness field in our model), \\
(iii)  adjacent block-faces have similar levels of occupancy. 
}

\section{Details pertaining to the Analytical Calculations}

\subsection{Search time in real time units \label{sec:Appendix_searchtime_real}}

In the main text, we have exposed the calculation of the search time expressed in arbitrary units (Eq.~\ref{eq:stime_alpha_spots}). To do so, drivers were split into distinct categories $\alpha$, each endowed with their own matrix of turning probabilities ${\uul{T}}\alp$ and generalised transition matrix $M\alp_{ij}= (1-p_i\alp\,\hat{n}_i)\cdot T_{ij}\alp$. Since ${M\alp}^K$ determines how the density of cars evolves on the graph in $K$ steps (i.e., hops from node to node), we were able to calculate the average number of steps before parking.

In order to derive a search time in minutes, a slightly more elaborate method is needed. For that purpose, we insert the transition matrix into a `generating' function
$z \rightarrow \Na(z)$, where $z$ is a real variable, $N\alp_{ij}(z)=z^{\tau_{ij}}\,M_{ij}\alp$, and $\tau_{ij}$ is the travel time from node $i$ to node $j$. At this stage, one can notice
that the exponentiation of this matrix to the power $K$ yields
\begin{equation}
\left[ \Na^K \right]_{ij} (z) = \sum_{\pi\,\mathrm{s.t.}\,\pi_0=0,\pi_K=j}z^{\tau_{\pi_{0}\pi_{1}}+\ldots+\tau_{\pi_{K-1}\pi_{K}}} \prod_{k=0}^{K-1}M_{\pi_{k}\pi_{k+1}},
\end{equation}
where the sum runs over permutations $\pi$ of indices (or `paths') such that $\pi_0=i$ and $\pi_K=j$. It immediately follows that
\begin{equation}
\frac{d}{dz}\left[ \Na^K \right]_{ij}(z=1)=\sum_{\pi\,\mathrm{s.t.}\,\pi_0=0,\pi_K=j} \left(\tau_{\pi_{0}\pi_{1}}+\ldots+\tau_{\pi_{K-1}\pi_{K}}\right) \prod_{k=0}^{K-1}M_{\pi_{k}\pi_{k+1}}
\end{equation}
is the probability to reach spot $j$ a number $K$ of steps after injection of the car
at node $i$, multiplied by the total travel time from $i$ to $j$. The (unbound)
travel time before parking thus reads
\begin{eqnarray*}
T_{s}\alp & = & H_i\alp(0) \cdot \sum_{K=0}^{\infty}
           \frac{d}{dz}\left[\Na^{K}\right]_{ij}(z=1)\cdot \tilde{p}\alp_j \\
 & = & H_i\alp(0) \cdot \frac{d}{dz}\left[ \left(\Id-\Na(z)\right)^{-1}\right]_{ij}(z=1) \cdot \tilde{p}\alp_j 
\end{eqnarray*}

But, since
$\frac{d\uul{A}^{-1}}{dz}=-\uul{A}^{-1}(z)\frac{d\uul{A}}{dz}\uul{A}^{-1}(z)$ for any
 differentiable function $z \rightarrow \uul{A}(z)$ of invertible matrices $\uul{A}(z)$
and $\Na(z=1)=\Ma$,
\begin{equation}
T_{s}\alp  =  H_i\alp(0) \cdot
\left[\left(\Id-\Ma\right)^{-1}
    \cdot \Na'(z=1)
    \cdot \left(\Id - \Ma \right)^{-1}
\right]_{ij}
\cdot  \tilde{p}\alp_j,
\label{eq:SI_T_search_realunits}
\end{equation}
where we recall that $N_{ij}^{(\alpha)\prime}(z=1)=\tau_{ij} M\alp_{ij}$ and $\tilde{p}\alp_j=\hat{n}_{j}p\alp_j$.

Equation~\ref{eq:SI_T_search_realunits} expresses the mean search time in minutes (or, more generally, in the same units as $\tau_{ij}$) as a function of the occupancy field ($n_j$).

\subsection{Capped search time in arbitrary and real time units \label{sec:app_capped_search} }

Equations~\ref{eq:P_j_alp} and \ref{eq:stime_alpha_spots} of the main text give the mean search time
of drivers that will hypothetically keep cruising forever. In reality, one expects them to quit searching after a given time. 

Let us first assume that this upper time bound is given as a number $K_{\max}$ of steps from node to node. Then, capping the search simply implies restricting the sum on $K$ in Eqs~\ref{eq:P_j_alp} and \ref{eq:stime_alpha_spots} to $0\leqslant K \leqslant K_{\max}$, which yields Eq.~\ref{eq:P_j_alp_capped} for the probability $\bar{P}_j\alp$ to reach spot $j$ and park there. The stationary occupancy field $(n_j)$ can then
be derived iteratively by inserting the capped reaching probabilities $\bar{R}_j\alp\equiv \bar{P}_j\alp / \tilde{p}_j$  into Eq.~\ref{eq:nj_stat_self}. 

Once the occupancy field is known, one can turn to the capped search time \AN{expressed as a number of steps},

\begin{eqnarray*}
\bar{\Tk}^{(\alpha)} & = & H_i\alp(0)\cdot\left[\sum_{K=0}^{K_{\max}}K\,\Ma^{K}+\sum_{K=K_{\max}+1}^{\infty}K_{\max}\,\Ma^{K}\right]_{ij} \tilde{p}\alp_j\\
 & = & H_i\alp(0)\cdot\left[\sum_{K=0}^{K_{\max}}K\,\Ma^{K}+K_{\max}\left(\Id-\Ma\right)^{-1}\cdot \Ma^{K_{\max}+1}\right]_{ij} \tilde{p}\alp_j\\
 & = & H_i\alp(0)\cdot\left[\sum_{K=0}^{K_{\max}}\sum_{l=0}^{K-1}\,\Ma^{K}+K_{\max}\left(\Id-\Ma\right)^{-1}\cdot \Ma^{K_{\max}+1}\right]_{ij} \tilde{p}\alp_j,
\end{eqnarray*}
where one has arbitrarily defined as  $K_{\max}$ the search time of cars that quit searching.

Using the following identity,
\begin{eqnarray*}
\sum_{K=0}^{K_{\max}}\sum_{l=0}^{K-1}\,\Ma^{K} & = & \sum_{l=0}^{K_{\max}-1}\sum_{K=l+1}^{K_{\max}}\,M^{K}\\
 & = & (\Id-\Ma)^{-1}\sum_{l=0}^{K-1}\Ma^{l+1}\left(\Id-\Ma^{K_{\max}-l}\right)\\
 & = & \Ma \cdot (\Id-\Ma)^{-2}\cdot (\Id-\Ma^{K_{\max}})-K_{\max} (\Id-\Ma)^{-1} \cdot \Ma^{K_{\max}+1} ,
\end{eqnarray*}
we arrive at 
\begin{eqnarray}
\bar{\Tk}^{(\alpha)} & = & H_i\alp(0)\cdot\left[ \Ma \cdot (\Id-\Ma)^{-2} \cdot (\Id-\Ma^{K_{\max}})\right]_{ij} \tilde{p}\alp_j \nonumber \\
 & = & \Tk\alp-H_i\alp(0)\cdot\left[ (\Id-\Ma)^{-2} \cdot \Ma^{K_{\max}+1}\right]_{ij} \tilde{p}\alp_j.
  \label{eq:SI_stime_spots_capped}
\end{eqnarray}
As in the main text, Einstein's summation convention on repeated indices is implied.

To recover real time units, we calculate an average conversion factor between steps $K$ and seconds using the case of unbound searches, by equating the (unbound) search time given by Eq.~\ref{eq:stime_alpha_spots} and that given by Eq.~\ref{eq:stime_alpha_real}. The maximum number of steps $K_{\max}$ is then first estimated from the maximum allowed time and the capped search time $\bar{\Tk}^{(\alpha)}$ is eventually converted into seconds on the same basis.

Unfortunately, evaluating the capped search time via Eq.~\ref{eq:SI_stime_spots_capped} involves the computation of $\Ma^{K_{\max}+1}$ for $K_{\max}\gg 1$, which will not necessarily be a sparse matrix even if $\Ma$ is. This computation turns out to be numerically very demanding if the number
of nodes in the graph is huge, as in the Lyon test case. However, the method is operational and quick on smaller networks. Consider for example the `toy' network introduced in Fig.~\ref{fig:smallLyon}(a). Capping search times to $180\,\mathrm{s}$ reduces the simulated mean search time all the more as the injected rate is high, as expected and shown on Fig.~\ref{fig:SI_search_times_smallLyon}. These capped times are very well captured
by the theoretical method outlined above, culminating in Eq.~\ref{eq:SI_stime_spots_capped}, as can be seen on  Fig.~\ref{fig:SI_search_times_smallLyon}.

\begin{figure}[!htb]
    \centering
        \includegraphics[width=0.62\textwidth]{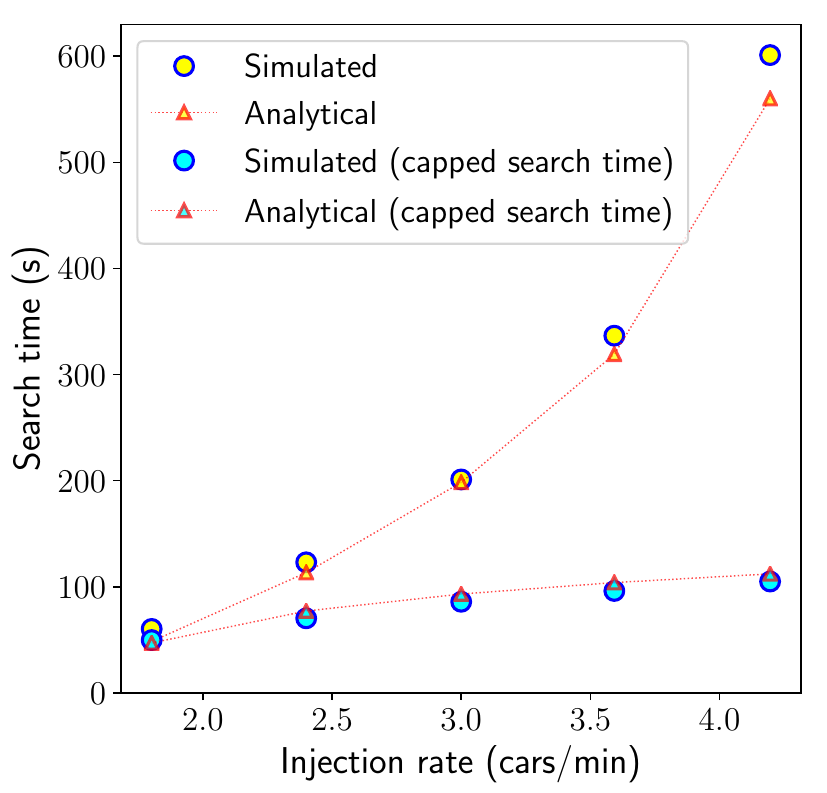}
\caption{Variations of the driving and cruising time in the `toy' network of Fig.~\ref{fig:smallLyon}(a) (with free spots) with the car injection rate. The outcome of the simulation (circles) is compared with the analytical predictions (triangles), both in the case of unbound search times and when search times are capped to $180\,\mathrm{s}$.}
\label{fig:SI_search_times_smallLyon}
\end{figure}

\section{Detailed Input for the Simulations of Lyon}

Section~\ref{sec:Lyon} presented a large-scale application of the proposed framework in the case of the city of Lyon, France. In this application, drivers are classified into 36 categories according to their
final destination (the list of which is given in Tab.~\ref{tab:SI_destinations}). Their cars are injected into the networks at one of the 49 entry points enumerated in Tab.~\ref{tab:SI_entries}, with a relative probability inferred either from the population of the surrounding neighbourhood or on a rough estimation of the inflow from the periphery of the city, if the entry point is located at its boundary.

\begin{table}[h]
\begin{tabular}[c]{|c|c|c|c|c|}
\hline 
Label & Name & $x$ & $y$ & Probability\tabularnewline
\hline 
\hline 
0&Saint-Rambert - industrie&841367.9&6523437.7&0.013\tabularnewline \hline
1&La Duchère&839095.5&6522460.5&0.009\tabularnewline \hline
2&Vaise&840269.9&6521523.7&0.047\tabularnewline \hline
3&Champvert – Point-du-jour&840501.8&6520877.9&0.016\tabularnewline \hline
4&Ménival - La Plaine&839660&6519395.7&0.012\tabularnewline \hline
5&Saint-Just&840845.5&6519201.4&0.012\tabularnewline \hline
6&Vieux-Lyon&841609.7&6519231.4&0.012\tabularnewline \hline
7&Chazière -Flammarion&841413.3&6520955.4&0.028\tabularnewline \hline
8&Cœur Croix-Rousse&842199.3&6521377.5&0.028\tabularnewline \hline
9&Les Chartreux&841921.4&6520617&0.018\tabularnewline \hline
10&Pentes&842580&6520652.1&0.013\tabularnewline \hline
11&Terreaux – Cordeliers&842555.7&6520004.7&0.019\tabularnewline \hline
12&Ainay&842195&6519007.4&0.026\tabularnewline \hline
13&Sud Perrache&841731.4&6517736.3&0.031\tabularnewline \hline
14&Tête d'or – Foch&843412.4&6520782.8&0.04\tabularnewline \hline
15&Brotteaux – Europe&844411.3&6520546.7&0.028\tabularnewline \hline
16&Bellecombe – Thiers&844989.5&6519947.4&0.041\tabularnewline \hline
17&Mutualité-Préfecture&843080.8&6519272&0.032\tabularnewline \hline
18&Part-Dieu - Bir Hakeim&843782.1&6519416.5&0.032\tabularnewline \hline
19&Paul Bert – Villette&845042.8&6519497.8&0.036\tabularnewline \hline
20&Dauphiné - Sans Souci&845321.7&6518826.4&0.032\tabularnewline \hline
21&Montchat &846350.6&6518432.2&0.032\tabularnewline \hline
22&Jean Macé&842992.7&6518754.3&0.037\tabularnewline \hline
23&Guillotière&843092.8&6518310.3&0.047\tabularnewline \hline
24&Blandan&843950.4&6518243.7&0.047\tabularnewline \hline
25&Gerland nord&842799.7&6517560.4&0.047\tabularnewline \hline
26&Gerland sud&842756.2&6516727.5&0.047\tabularnewline \hline
27&Grand Trou – Moulin à vent&844554.8&6516452.7&0.027\tabularnewline \hline
28&Monplaisir&844588.2&6517391.2&0.027\tabularnewline \hline
29&Le Bachut&844966.5&6517711.6&0.036\tabularnewline \hline
30&Etats-Unis&845574.1&6517087.7&0.023\tabularnewline \hline
31&Mermoz – Laennec&846128.7&6516429&0.022\tabularnewline \hline
32&Général André -  Santy&846210.7&6515824.3&0.022\tabularnewline \hline
33&Pinel&845979&6518623&0.014\tabularnewline \hline
34&6e arrondissement Sud&843836.9&6520125.1&0.040\tabularnewline \hline
35&5e arrondissement Sud&838846.3&6518189.5&0.012\tabularnewline \hline
\end{tabular}

\caption{\label{tab:SI_destinations}List of the 36 destinations implemented in our study of Lyon. The `probability' column specifies the fraction of cars bound to a given destination. Coordinates are given in the RGF-93/Lambert-93 reference system.}
\end{table}

\clearpage

\begin{table}[H]
\begin{longtable}[c]{|c|c|c|c|}
\hline 
Label  & $x$ & $y$ & Proba of injection $I_i(0)$\tabularnewline
\hline 
\endhead
\hline 
0&843661&6520685&0.017\tabularnewline \hline
1&841446&6516539&0.161\tabularnewline \hline
2&844237&6520113&0.017\tabularnewline \hline
3&840046&6520488&0.024\tabularnewline \hline
4&842098&6515492&0.011\tabularnewline \hline
5&846715&6516345&0.048\tabularnewline \hline
6&846268&6515772&0.012\tabularnewline \hline
7&843023&6521479&0.017\tabularnewline \hline
8&841709&6517901&0.014\tabularnewline \hline
9&844843&6520013&0.017\tabularnewline \hline
10&846215&6518226&0.020\tabularnewline \hline
11&843218&6515591&0.011\tabularnewline \hline
12&844862&6519680&0.020\tabularnewline \hline
13&842529&6520514&0.007\tabularnewline \hline
14&844989&6517334&0.012\tabularnewline \hline
15&839541&6521161&0.064\tabularnewline \hline
17&842670&6517036&0.011\tabularnewline \hline
18&844226&6518050&0.011\tabularnewline \hline
20&841862&6516427&0.011\tabularnewline \hline
21&841071&6522242&0.017\tabularnewline \hline
22&847245&6517797&0.039\tabularnewline \hline
23&841429&6520673&0.007\tabularnewline \hline
24&844289&6516742&0.012\tabularnewline \hline
25&842837&6518111&0.011\tabularnewline \hline
26&842041&6520903&0.007\tabularnewline \hline
27&839305&6519151&0.012\tabularnewline \hline
28&839305&6519151&0.012\tabularnewline \hline
29&841647&6520269&0.012\tabularnewline \hline
30&845762&6518982&0.023\tabularnewline \hline
31&839240&6521852&0.024\tabularnewline \hline
32&846693&6515848&0.012\tabularnewline \hline
33&844198&6515464&0.048\tabularnewline \hline
34&843204&6518705&0.011\tabularnewline \hline
35&844645&6516244&0.048\tabularnewline \hline
36&840767&6519147&0.012\tabularnewline \hline
38&845656&6516640&0.012\tabularnewline \hline
39&843257&6519411&0.020\tabularnewline \hline
40&844439&6520869&0.042\tabularnewline \hline
41&844554&6517528&0.012\tabularnewline \hline
42&844889&6516878&0.012\tabularnewline \hline
43&842446&6521030&0.007\tabularnewline \hline
44&844431&6519483&0.020\tabularnewline \hline
45&845021&6519236&0.020\tabularnewline \hline
47&842325&6519031&0.015\tabularnewline \hline
48&839654&6522705&0.023\tabularnewline \hline

\end{longtable}

\caption{\label{tab:SI_entries}List of the 49 entry points considered for the injection of cars into the street network, with their relative probabilities. Coordinates are given in the RGF-93/Lambert-93 reference system.}
\end{table}

\clearpage

 \SingleSpacedXII
 
\section{Numerical Implementation of the Agent-based Model \label{sec:app_implementation}}

 The numerical implementation of the agent-based model relies on the C++ classes: 
 
  \textbf{\underline{City}} \\
  \emph{Attributes}: name, number of car categories, list of streets, list of nodes, matrices of turn-choices for all car categories, list of active cars, list of parked cars, various storage vectors

   \textbf{\underline{Street}} \\
    \emph{Attributes}: name, identifier, start and end nodes, number of spots, length, effective speed, various storage vectors
   
    \textbf{\underline{Node}} \\
     \emph{Attributes}: identifier, GPS coordinates, lists of incoming and outgoing streets
    
     \textbf{\underline{Spot}} \\
      \emph{Attributes}: (inherits from Node), identifier, vacancy status, attractiveness, various storage vectors

 \textbf{\underline{Car}} \\
 \emph{Attributes}: car number, car type, car status (active, searching for parking, parked, frozen, exited), destination, beta, position (on street or spot), matrix of turn-choices, various timestamps.
 
 The main loop of the script runs as explained  in Alg.~\ref{alg:mainloop} in pseudo-code:

 \begin{algorithm}[h]

\caption{Main loop}
 \label{alg:mainloop}
\begin{algorithmic}
\FOR{t in all time steps ($dt=1\,\mathrm{s}$)}
\STATE
    - draw number of new cars to be injected from a Poisson distribution
    \FOR{car in cars to be injected}
    \STATE
       - compute the normalized cumulative distribution function (cdf) of all probabilities at all possible injection nodes \\
        -  select an injection node among the possibilities by mapping a random number onto the interval of the computed cdf \\
        - inject the car at the chosen node
    \ENDFOR

    \FOR{car in active cars}
    \STATE
         - update the perceived tension $\beta$
         \WHILE{current timestep is not over}
         \STATE
            - move the car so far as possible along current street\\
            - store in a list $L_\mathrm{spots}$ the spots by which the car has driven\\
            - compute cumulative distribution function (cdf) of probabilities of all possible parking choices in $L_\mathrm{spots}$ based on attractiveness\\
            \FOR{spot in $L_\mathrm{spots}$}
            \STATE
                - decide if car parks here by mapping a random number on the interval of the computed cdf\\
                \IF{car parks here}
                \STATE
                    - end `WHILE' loop
                \ENDIF
            \ENDFOR
            \IF{car has reached end of street}
            \STATE
                    - compute the cdf of probabilities of all possible transitions available from the node at the end of street\\
                    - select the outgoing street in which to move the car in the usual way\\
                    - move the car into next street
            \ENDIF
            \STATE
            - record car position\\
            - update leftover time in current timestep
         \ENDWHILE
    \ENDFOR
    \FOR{car in parked cars}
    \STATE
        - compute departure rate\\
        - randomly decide if car is removed (departs)
    \ENDFOR\\
    
    - keep track of the global balance of injected, active, parked and departed cars
\ENDFOR

\end{algorithmic}
\end{algorithm}

 \end{APPENDICES}

\end{document}